\documentclass[aps,prd,twocolumn,groupedaddress]{revtex4}
\usepackage[dvips]{graphicx,color} 
\usepackage{array,hhline,multirow,dcolumn} 
\usepackage{rotating} 
\usepackage{amsmath}
\usepackage{amsfonts}
\usepackage{amssymb}
\usepackage{graphicx}%
\usepackage{graphics}%

\begin{document}


\title{Precision Measurement of $\sin^2 \theta_W$ at a Reactor}


\author{J.~M.~Conrad}
\author{J.~M.~Link}
\author{M.~H.~Shaevitz}
\affiliation{Columbia University, Dept. of Physics, New York, NY 10027, USA}


\date{\today}

\begin{abstract}
This paper presents a strategy for measuring $\sin^2 \theta_W$ to
$\sim$1\% at a reactor-based experiment, using $\overline{\nu} e$ 
elastic scattering.  This error is comparable to the NuTeV, SLAC
E158, and APV results on $\sin^2 \theta_W$, but with substantially
different systematic contributions.  The
measurement can be performed using the near detector of the
presently proposed reactor-based oscillation experiments.  We
conclude that an absolute error of $\sim\!\delta({\mathrm{sin}}^2
\theta_W)=0.0019$ may be achieved.
\end{abstract}

\pacs{12.15.Ji,13.15.+g,14.60.Lm,28.50.Hw}

\maketitle

This paper outlines a method for measuring $\sin^2 \theta_W$
$(Q^2\approx0)$ at a reactor-based experiment.  The study is motivated
by the NuTeV result, a $3 \sigma$ deviation of $\sin^2 \theta_W$ from
the Standard Model prediction~\cite{NuTeV}, measured in deep inelastic
neutrino scattering ($Q^2 =$ 1 to 140 GeV$^2$, $\langle Q^2_\nu
\rangle=26$ GeV$^2$, $\langle Q^2_{\overline{\nu}} \rangle =15$ GeV$^2$).
Various Beyond-the-Standard Model explanations have been put
forward~\cite{sampage,Davidson,Loinaz:2002ep}, and, to fully resolve the 
issue, many require a follow-up experiment which probes the
neutral weak couplings specifically with neutrinos, such as the one
described here.

\begin{figure}
\vspace{5mm}
\centering
\includegraphics[width=7.0cm, bb=32 39 475 665, clip=true]{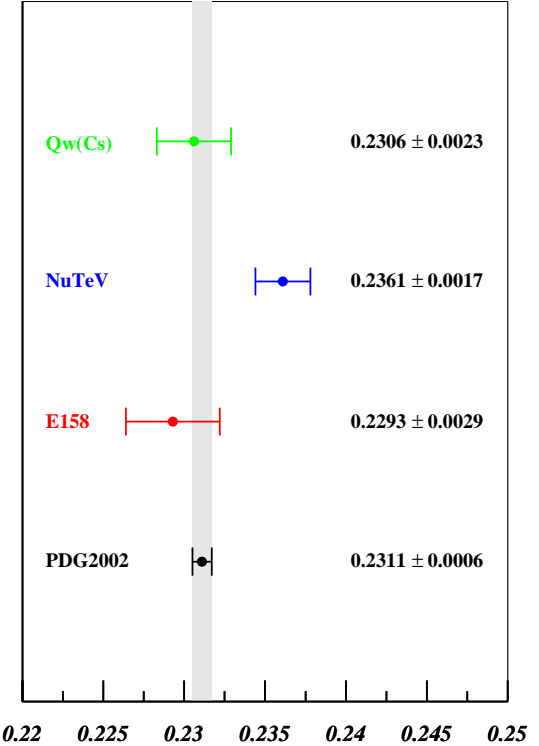}
\vspace{1mm}
\caption{\label{Marciano}Measurements of $\sin^2 \theta_W$, extrapolated to the $Z^0$ mass, from reference~\cite{Anthony:2003ub}.}
\end{figure}

This proposed measurement is also interesting as an additional
precision study at $Q^2=4\!\times\!10^{-6}$ GeV$^2$.  The two existing low
$Q^2$ measurements are from atomic parity violation (APV)~\cite{APV}, which
samples $Q^2 \sim 10^{-10}$~GeV$^2$; and SLAC E158, a M{\o}ller scattering experiment
at average $Q^2=0.026$ GeV$^2$~\cite{Anthony:2003ub}.  Using the measurements at
the $Z$-pole with $Q^2=M_z^2$ to fix the value of $\sin^2 \theta_W$,
and evolving to low $Q^2$, Fig.~\ref{Marciano} shows that these
results are in agreement with the Standard Model.  However, the
radiative corrections to neutrino interactions allow sensitivity to
high mass particles which are complementary to the APV and M{\o}ller
scattering corrections.  Thus, this proposed measurement will provide
valuable additional information.
 
The technique we employ uses the rate of the purely leptonic $\overline{\nu}
e$ scattering to measure $\sin^2 \theta_W$.  This signal was first
detected by Reines, Gurr, and Sobel~\cite{reines}, who measured $\sin^2
\theta_W = 0.29\pm 0.05$.  In this paper, we explore what is necessary
to improve on their idea and make a competitive measurement today.
One important step is to normalize the $\overline{\nu} e$ ``elastic
scatters'' using the $\overline{\nu} p$ inverse beta decay events, to
reduce the error on the flux.  Other crucial improvements are that the
detector is located beneath an overburden of $\sim$300 mwe (meters,
water-equivalent) and built in a clean environment.  We find that a
measurement of $\pm 0.0020$ is achievable.  This is comparable to the
NuTeV error of $\pm 0.00164$, and may help clarify the 
theoretical situation, as shown in references~\cite{Rosner:2004yt,Fisher}.

The proposed design employs spherical scintillator oil detectors
similar to those used by CHOOZ~\cite{CHOOZfinal} and by other
experiments which have been proposed to measure the oscillation parameter
$\theta_{13}$~\cite{whitepaper}. 

This style of detector has been optimized to reconstruct $\overline{\nu} 
p\to e^+ n$ events, which dominate the rate when the reactor is running.  
We show that this design is also suitable for measuring $\sin^2
\theta_W$ to high precision.  Initially, one might think otherwise,
since the $\overline{\nu} p$ events represent a potential background.  
However, this background can be controlled.  In fact, these events are 
invaluable because they provide the normalization constraint.
This normalization measurement must be done in the same detector as
the $\overline{\nu} e$ measurement to exploit cancellations of systematics,
especially those related to the fiducial volume.

To control backgrounds, this analysis exploits a visible
energy ($E_{vis}$) ``window''.  We will show that we can obtain
significant signal statistics even in this limited region.  On the
other hand, this range is above most environmental backgrounds in the
detector, and below the energy produced by neutron capture in Gd.

This paper is organized in the following manner.  In Sec.~\ref{introdesign}, 
we identify the important questions which drive the design choices.  In 
Sec.~\ref{generic}, we provide details of the generic experiment and analysis 
used for estimates.  In Sec.~\ref{nubare}, we discuss $\overline{\nu} e$ 
event identification and rates.  In Sec.~\ref{nubarp}, we discuss rejection 
of $\overline{\nu} p$ events.  In Sec.~\ref{environ}, we consider backgrounds 
produced by natural radioactivity and cosmic rays.  In Sec~\ref{norm}, we 
consider the errors on the $\overline{\nu} p$ normalization sample.  In 
Sec.~\ref{errors}, we discuss how we find the error on $\sin^2\theta_W$.  
Finally, in Sec.~\ref{conclusions}, we present our conclusions.

The goal of this paper is to establish that this analysis is worth pursuing 
at a reactor-based experiment.  Thus the analysis is presented in
sufficient detail to address what we have identified as the major issues.  
Many detailed studies remain to be done, however, as we discuss in the 
conclusions.  In order to demonstrate feasibility we have relied on 
techniques for reducing background which are well-established in our 
determination of the error on $\sin^2 \theta_W$. 

\section{\label{introdesign}Introduction to the Design Issues}

NuTeV has made a 0.72\% measurement, including statistics and
systematics, of the weak mixing angle.  This error corresponds to 
a 1.15\% uncertainty on the $\overline{\nu} e$ absolute rate
at a reactor experiment. With this in mind, in order to establish the
design for this experiment, the following questions must be explored:
\begin{enumerate}
\item Are there sufficient elastic scattering events to perform this
  measurement?

\item Can the elastic scattering events be isolated from the inverse 
beta decay events?

\item Can the environmental backgrounds be controlled?

\item How well can the anti-neutrino flux be known?

\end{enumerate}
This section provides qualitative answers to establish that an error
on $\sin^2 \theta_W$ comparable to NuTeV is feasible.

This section also provides simple motivations for the major cuts.
Briefly sketched, these are: a fiducial volume cut which is well
within the Gd-doped region; vetoes for cosmics; an energy window cut; 
and a timing window to search for neutrons which follow a
neutrino interaction.  Here, we aim only to address the basic needs
and challenges.  The specifics on the cuts are described in
Sec.~\ref{cuts}.  The consequences of the cuts are explored in
Sec.s~\ref{nubare} through \ref{norm}.

Throughout the paper, we will identify certain backgrounds as
``negligible.''  We define negligible as a contribution to the total
error of $\le 0.1$\%.  In the case of backgrounds to the signal, this
amounts to $<10$ events.

\subsection{Statistics}

\begin{figure}
\vspace{5mm}
\centering
\includegraphics[width=8.5cm,bb=25 416 528 646,clip=true]{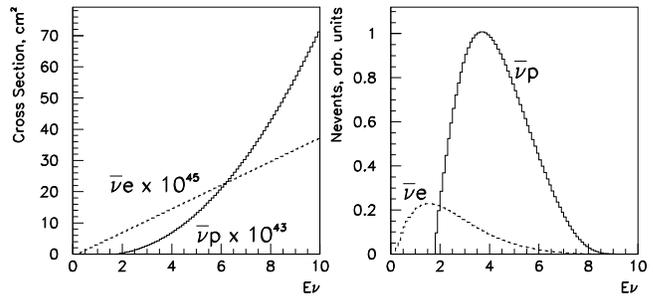}
\vspace{1mm}
\caption{\label{fig:xsec} Left:  Comparison of $\overline{\nu} e$ (Elastic Scatters) and $\overline{\nu} p$ (inverse beta-decay) cross sections as a function of neutrino energy in MeV.  Right: Comparison of event rates for $\overline{\nu} e$ and $\overline{\nu} p$ as a function of neutrino energy in MeV.  Note that electron targets exceed free proton targets in the oil by a factor of 4.3.}
\end{figure}

The signal sample consists of elastic scattering events (``Elastic 
Sample'').  A $\le 1$\% statistical error, corresponding to 
$\ge$10,000 elastic events, is necessary if the goal is a total error 
comparable to NuTeV.

\begin{figure}
\vspace{5mm}
\centering
\includegraphics[width=8.5cm,bb=72 370 540 654,clip=true]{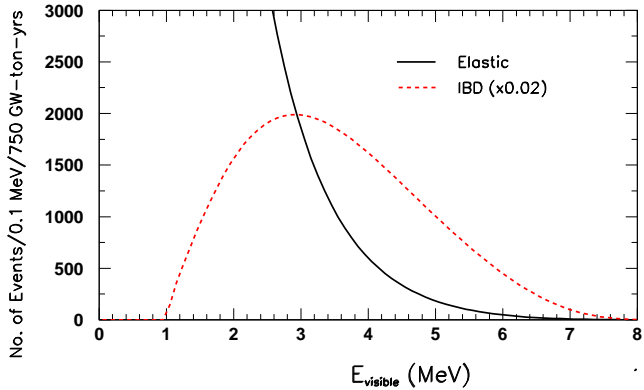}
\vspace{1mm}
\caption{\label{fig:visiblecomp}Visible energy distribution for inverse beta decay (red, dashed) and elastic scattering events (black, solid).  Inverse beta decay events are scaled by a factor of 0.02 to allow visual comparison.}
\end{figure}

The number of elastic scattering events and inverse beta decay events
scale together.  Since our premise is to use near detectors for the
$\theta_{13}$ measurement, which utilizes inverse beta decay events,
it is instructive to understand the relative rates of these two
processes.  Fig.~\ref{fig:xsec} (left) compares the cross sections for
these interactions as a function of neutrino energy in MeV, scaled for
convenience.  Fig.~\ref{fig:xsec} (right) compares the unscaled number
of events.  At low energies, where
$\overline{\nu} p$ is kinematically suppressed, elastic scattering
dominates.  Finally, Fig.~\ref{fig:visiblecomp} compares the absolute
event rates as a function of visible energy in the detector.  The
elastic scattering events peak at low visible energy due to the energy
carried away by the outgoing neutrino.  
From Fig.~\ref{fig:visiblecomp}, one can see that if greater than $1\!\times\!
10^6$ $\overline{\nu} p$ events can be collected in the visible energy
window, then one will obtain more than $1\!\times\!10^5$ $\overline{\nu} e$
events.  Thus the necessary statistical precision of $<1$\% on elastic
scattering can be reached.

Based on this, we require a design which results in $>1\!\times\!10^6$
$\overline{\nu} p$ events.  This goal is in concert with the requirements
for a near detector for a $\theta_{13}$ measurement~\cite{whitepaper}.
The designs under consideration build on the past experience of the
CHOOZ experiment, which observed $\sim$3000~inverse beta decay events
in a 5~ton detector located 1~km from two 4.5~GW reactors, running for
132~days effective full power~\cite{CHOOZfinal}.  The proposed near
detectors are typically located about 200~m from the reactor, gaining
a factor of 25 from solid angle.  The detector will be built with
increased fiducial mass.  Multiple detectors can be built.  The
experiment can feasibly run longer.  In summary, the necessary event
rate appears to be attainable with reasonable modifications to the
CHOOZ setup.

\subsection{$\overline{\nu} p$ Mis-identification}

Inverse beta-decay events are a major component of the reactor-on rate
in the proposed visible energy window. The best method for separating
these events from elastic scatters is observation of the signal from
neutron capture.  This will motivate a fiducial volume cut which is
well within the Gd-doped region to assure high efficiency for
capturing the neutron.  It will also motivate a data acquisition system which is
sensitive to neutron capture on H, which occurs 16\% of the time
despite the Gd doping.  Lastly, it will motivate an efficient
time window for the neutron search.  These are all discussed
in detail in Sec.~\ref{nubarp}.

\subsection{\label{intro_env}Environmental Backgrounds}

\begin{figure}
\vspace{5mm}
\begin{center}
\includegraphics[width=8.5cm,clip=true]{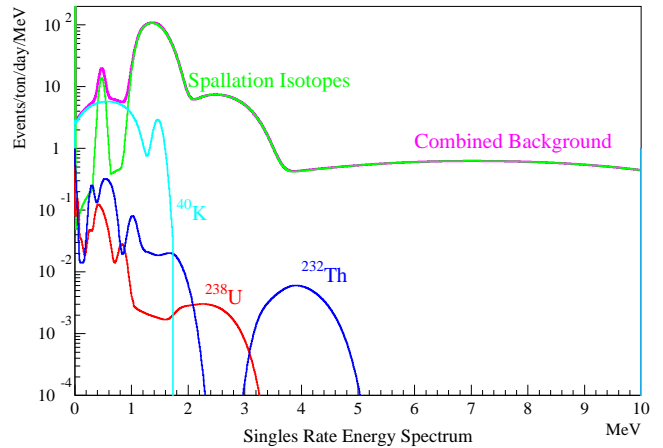}
\vspace{1mm}
\caption{\label{fig:UTh}Simulated visible energy distribution and sources of singles events in a scintillation detector located at a depth of 300~mwe with a flat overburdon.  The assumed isotope concentrations come from KamLAND~\cite{Nakamura:2004cb}.}
\end{center}
\end{figure}

Environmental backgrounds are by far the most important issue in the
analysis and therefore deserve substantial introduction here.  They
fall into two categories: naturally occurring radioactivity and
muon-induced backgrounds.  To gain a sense of the background expectations,
Fig.~\ref{fig:UTh} shows a simulated visible energy spectrum for singles 
events consistent with Braidwood~\cite{braidwood}.  The simulation 
assumes complete containment of all decay daughters, and uses an energy 
resolution of 7.5\%/$\sqrt{E(MeV)}$, and  a Birks' constant of 
0.00165~cm/MeV to quench the energy response to alphas.  The naturally 
occurring radioactive contaminants mainly populate the low energy range 
of Fig.~\ref{fig:UTh} and can, in part, be kept under control by 
maintaining KamLAND standards of oil purity, but, unlike KamLAND, this 
experiment will use Gd-doped scintillator, and so the Gd must also be 
purified of radioactive contaminants.  The other source of environmental 
background, the $\beta$-decays of muon-induced (or ``spallation'') 
isotopes, populate the higher energy range of Fig.~\ref{fig:UTh}.

To reduce background from radioactivity, we introduce spatial and
energy cuts.  Activity from the tank walls, the phototubes, and the
acrylic vessel separating the Gd-doped and undoped regions can be
removed by a strong fiducial volume cut.  Most background from 
radioactivity dissolved in the scintillator can be removed from the 
sample through a $3<E_{vis}<5$~MeV cut on visible energy, as seen in 
Fig.~\ref{fig:UTh}.  However,the $^{232}$Th chain produces $^{208}$Tl, 
which $\beta$-decays in our visible energy window and must be addressed.

Potential background from cosmic rays comes from 1) the muons
themselves; 2) electrons from muon decays (``Michel electrons''); 3)
$^{12}$B decays from $\mu^-$ capture; 4) spallation neutrons; and 5)
isotopes generated by the high energy muons.  The first four are
straightforward to reduce.  The fifth is, potentially, the most
significant background in this analysis.

First, consider the four which are straightforward.  Muons which enter
the tank can be easily identified by means of the large energies which
they deposit.  Muons which stop may decay to produce electrons, or
capture to produce $^{12}$B, which $\beta$-decays. The need to
identify stopping muons motivates a veto based on a combination of
tank hits and lack of hits in a hodoscope below the tank.  Neutrons
which are produced in combination with a cosmic ray event will be
identifiable by their capture.  Spallation neutrons which are
unassociated with a cosmic ray have two sources.  They may be produced
outside of the tank and then enter; or they may be produced by high
energy muon interactions with the $^{12}$C in the tank, but not be
associated with the parent cosmic due to a late capture time.  The
Gd-doped buffer region surrounding the fiducial region provides
further protection from incoming neutrons.  We will show that the
proposed visible energy window eliminates unassociated neutrons in
the tank.

\begin{table}[tbp]
\begin{center}
\begin{tabular}{|c|c|c|c|c|} \hline
Isotope  &  Source \\ \hline
$^{9}$Li & $^{12}{\rm C} + \mu \rightarrow 3{\rm p} + ^{9}{\rm Li}+ \mu$ \\
$^{8}$He & $^12{\rm C} + \mu^- \rightarrow {\rm d} + {\rm 2p} + ^{ 8}{\rm He}$; 
$^{12}{\rm C} + \mu \rightarrow  {\rm 4p} + ^{8}{\rm He} +\mu$\\
$^{8}$Li & $^{12}{\rm C} + \mu \rightarrow  {\rm 3p} + {\rm 1n}+ ^{8}{\rm Li}+ \mu$\\ 
$^{6}$He & $^{12}{\rm C} + \mu \rightarrow \alpha + {\rm 2p} +  ^{6}{\rm He} + \mu$\\ 
$^{9}$C &  $^{12}{\rm C} + \mu \rightarrow {\rm 3n} +  ^{9}{\rm C} + \mu$\\ 
$^{8}$B & $^{12}{\rm C} + \mu \rightarrow {\rm 3n} + {\rm 1p} +  ^{8}{\rm B} + \mu$\\ 
$^{12}$B & $^{12}{\rm C} + n \rightarrow p + ^{12}{\rm B}$ \\ \hline 
\end{tabular}
\caption{\label{isotopesource}Examples of sources of isotopes which $\beta$ decay producing potential background to this analysis. }
\end{center}
\end{table}

Production mechanisms for the fifth source, $\beta$-decaying isotopes
produced by high energy muons, are listed in
Table~\ref{isotopesource}.  These isotopes are the dominate contribution 
to the singles rate above 3~MeV, as indicated by the ``Spallation 
Isotope'' curve in Fig.~\ref{fig:UTh}.  These are $^{9}$Li, $^{8}$He, 
$^{8}$Li , $^{9}$C , $^{8}$B, $^{12}$B, all of which have endpoints 
above 10~MeV; and $^{6}$He, which has an endpoint of 3.5~MeV.  $^{11}$Be 
is not considered in the standard
analysis because its muon induced production has only been reported 
as an upper limit~\cite{Hagner}.  However, we do consider the case where 
this contribution is equal to the limit as an alternate scenario in Sec.~\ref{errors}.

The most important and straightforward way to reduce the rates of
these isotopes is to have a large overburden.  In addition a 
veto system is also employed.  The
veto system must be more elaborate than a simple rejection of events
following an incoming cosmic ray, because the long half-lifes of the
isotopes results in an intolerable deadtime with this configuration.
However, a veto which identifies the subset of parent cosmics with evidence of an accompanying hadronic shower results in a tolerable deadtime.
This is called a muon-hadron veto, and is described in
Sec.~\ref{splat}.  This veto is the only proposed cut which is not
based on past experience.

\subsection{Normalization}

Absolute knowledge of the reactor neutrino flux is limited to
$\sim$2\% due to uncertainties on the reactor power and fuel composition.  
To avoid this systematic, we use $\overline{\nu} p$ events (the ``Normalization 
Sample'') to establish the normalization for the $\overline{\nu} e$ events.  
The statistical error on the $\overline{\nu} p$ events is small since more than 
$1\!\times\!10^6$ events are expected.  The cross section for $\overline{\nu} 
p$ is well known from theory, as discussed in Sec.~\ref{nubarp}, so
the systematic error from this source is negligible.  An important
systematic error comes from determination of the ratio of targets for
$\overline{\nu} e$ versus $\overline{\nu} p$ scatters, {\it i.e.} the
electron-to-free-proton ratio.  Another important systematic question
is related to neutron identification.  One can obtain a very pure
sample of $\overline{\nu} p$ events by requiring a Gd capture.  This,
however, will introduce a systematic error from the ratio of Gd
captures to the total.  This error was 1\% in CHOOZ.  This is unacceptably
high for this analysis and must be reduced through improved
calibration studies.  Alternatively, assuming the trigger has high
efficiency for events with H captures, one can accept all
$n$-identified events into the $\overline{\nu} p$ sample.  This eliminates
the error on the Gd capture ratio but introduces possible backgrounds
from accidental coincidences.  Estimating these backgrounds will
require a detailed study, beyond the scope of the present work.
Therefore, for this analysis, we will use the former method of
identifying a clean sample through the Gd captures.

\section{\label{generic}The General Design and ``Standard'' Analysis Cuts}

\begin{table}[tbp]
\begin{center}
\begin{tabular}{|c|c|} \hline
Days of running: &  900 Days \\ 
Number of reactor cores: & 2 \\
Power of each core:  & 3.6 GW \\
Overburden: & 300 mwe \\
Distance to near detectors: & 224 m \\ 
Number of near detectors: & 2 \\
Number of far detectors: & 4 \\ \hline
\end{tabular}
\caption{\label{overview}Overview of general assumptions}
\end{center}
\end{table}

To calculate an expected error on $\sin^2 \theta_W$, we must
make assumptions about the design.  A summary of the assumptions is
presented in Table~\ref{overview}.  The setup is drawn from an 
preliminary design of the Braidwood $\theta_{13}$ 
experiment~\cite{braidwood}.  The site has two 3.6~GW reactors which are 
assumed to produce a neutrino flux consistent with 
reference~\cite{GrattaRMP}.  The model for this study uses two near 
detectors, located 224~m from the reactors, and four far, which are located
1.8~km away (here the far detectors are used only to measure backgrounds). 
All six  detectors are assumed to be identical spherical vessels with 
both active and passive shielding.  Data taking is assumed to extend over 
900 live-days.

\begin{table}[tbp]
\begin{center}
\begin{tabular}{|c|c|}
\hline
\multicolumn{2}{|c|}{{\bf Basic Detector Design Parameters}}\\ \hline \hline
Radius of fiducial region        & 150 cm\\
Fiducial volume per detector     & 13 tons \\
Outer radius of central region   & 190 cm \\
Tonnage of the central region    & 26.5 tons \\
Outer radius of photon catcher   & 220 cm \\
Outer radius of detector         & 290 cm \\ \hline \hline
\multicolumn{2}{|c|}{{\bf Path lengths of Particles (for Containment)}}\\ \hline \hline
e$^-$ and e$^+$ track length      & negligible \\
e$^+$ to n separation length (for $\overline{\nu}p$ events)     & 6 cm \\
0.5 MeV $\gamma$ Compton path length & 11 cm \\ \hline \hline
\multicolumn{2}{|c|}{{\bf Neutron Parameters (for ID Efficiency)}}\\ \hline \hline
Fraction of n which capture on Gd (H) & 84\%  (16\%) \\ 
Neutron capture time & 30.5~$\mu$s \\  \hline
\end{tabular}
\caption{\label{tab:detassump}Assumptions related to the detector design used in this paper.}
\end{center}
\end{table}

It is necessary to make some specific assumptions in order to proceed
with our calculations.  These choices are reasonable and so serve for
the proof-of-principle calculation.  Small variations of this
``generic'' plan are expected and can easily be accommodated.
Table~\ref{tab:detassump} summarizes the assumptions.  We address 1)
basic definition of detector regions, 2) assumptions about track
length which are relevant to calculating backgrounds, and 3) parameters
related to identification efficiency which are relevant for
calculating both backgrounds and the normalization rate.

\subsection{The Basic Detector Design}

The outer radius of the detector design is chosen to allow the
detectors to fit in a 3~m radius tunnel. 
The interior sizes are scaled to match this
requirement. The detector has a ``central region'' of Gd-doped
scintillator, a ``photon catcher'' region which surrounds this, and an
``oil buffer'' region which separates the active regions from the
phototubes and tank walls.  For the sake of this discussion, we take
the outer radius of the central region to be 190 cm.  The fiducial
region must be of substantially smaller radius to maximize containment
of the neutrons produced by $\overline{\nu} p$ events and minimize
environmental backgrounds.  We will assume a fiducial radius of 150~cm.  
The Gd-doped region is surrounded by a 30 cm ``photon catcher''
of scintillator with no Gd doping.  The photon catcher permits high
efficiency for observing the 0.5~MeV $\gamma$'s produced by
annihilation in $\overline{\nu} p$ events.  These two regions are, in turn,
surrounded by an oil ``buffer'' in which the phototubes are immersed.
The buffer region extends out to a 290~cm radius.  Hence the buffer is 70~cm
in thickness and phototubes are located about 100~cm from the central
region or 140 cm from the fiducial region.

Given this layout, the fiducial volume of each detector contains 13 tons.
Therefore, two near detectors are required to attain the necessary 
statistics.  This is consistent, for example, with the Braidwood Experiment design~\cite{braidwood}.

\subsubsection{\label{reactorevents}Response to Reactor-induced Events}

The goal of the detector is to identify and count the two types of
reactor-induced events: elastic scattering and inverse beta decay.  In
order to do this, accurate energy and vertex reconstruction are
required.  Also, it is necessary to identify neutrons produced in
inverse beta decay with high efficiency.  It is worth noting that, for
this analysis, it is not necessary to reconstruct the angle of the
outgoing lepton.  This is in keeping with the detector design, where
the the high level of scintillation light will obscure any directional
Cerenkov light.

The two types of events have different visible energy distributions
(Fig.~\ref{fig:visiblecomp}),  so to relate the rates for the 
two processes, one needs a good understanding of the energy resolution 
of the detector.  Based on previous experiments, an energy resolution of $\le$10\% appears to be attainable~\cite{CHOOZfinal,CTFres,kamlandprl}.  
In Sec.~\ref{norm}, we show that systematics on smearing due to energy
resolution leads to a negligible systematic error in the analysis whereas the uncertainty on energy calibration is significant.

\begin{figure}
\vspace{5mm}
\centering
\includegraphics[width=8.5cm,bb=60 370 538 654,clip=true]{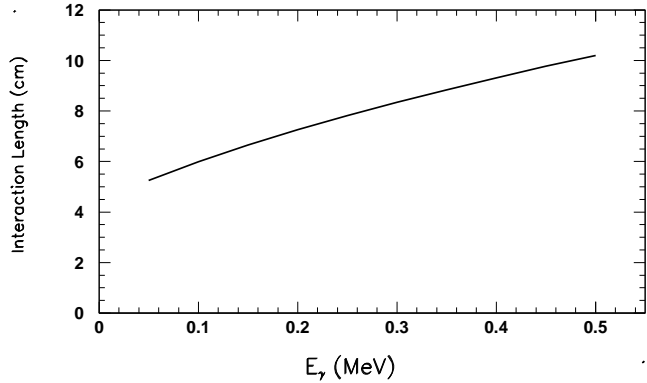}
\vspace{1mm}
\caption{\label{complen}Compton length in CH$_2$ as a function of photon energy.}
\end{figure}

To obtain good energy resolution for the normalization sample, the
annihilation photons in inverse beta decay events must be contained.
These photons lose energy through Compton scatters, with a path length
that depends on energy and in CH$_2$, which is $\sim$11~cm at 0.5~MeV (see
Fig.~\ref{complen}).  While the Compton peak is at $2/3 E_\gamma$,
note that the average energy loss is $1/3 E_\gamma$.  Thus an event
can be expected to have several Compton scatters before exiting the
detector.  The ``photon catcher'' region and outer 40~cm of the
Gd-doped region are used to contain and reconstruct the energy of
these photons.  The 1.5~m radius fiducial volume cut places a 0.5~MeV
photon at approximately 6.5 path lengths from the inactive oil-buffer
region.   The result is negligible loss due to escaping photons.

We do not consider vertex resolution smearing at the edge of the fiducial 
region.  Instead, we assume that the relative vertex resolution is
the same for the $\overline{\nu} e$ signal and $\overline{\nu} p$ normalization
events to within 1~mm.  Hence, the systematics related to vertex
resolution cancel.  We note that good vertex resolution is
important for identifying and removing backgrounds.  We believe that
$\sim$4~cm on the interaction vertex may be attainable.  This is
consistent with CHOOZ laser flasher studies~\cite{CHOOZfinal}.
CTF obtained a similar resolution using an alpha source, corresponding
to a photon energy of 0.862~MeV~\cite{CTFres}.  This is sufficiently
good that we believe vertex resolution issues will be a small effect
in the final analysis and they are not considered further here.

To identify $\overline{\nu} p$ events, which represent both a
background and a normalization sample, the signal from the neutron
capture is used.  In Gd-doped scintillator, a typical separation
length from neutrino vertex to neutron capture is 6 cm, as measured in
CHOOZ~\cite{CHOOZfinal}.  Therefore, the fiducial volume cut of 40 cm
from the central region edge represents 6.7 separation lengths.  Only
a small fraction of the neutrons are produced at the edge of the fiducial
region, and, of those, only about half drift outward.  Folding in the
geometry, assuming a uniform distribution for neutron production
throughout the tank, $1.1\!\times\!10^{-5}$ of the neutrons will exit without
capturing.  This contribution to the background will be considered
further in Sec.~\ref{nubarp}.

Neutron capture is delayed with respect to the positron track, with a mean 
capture time of 30.5~$\mu$s (as measured by CHOOZ~\cite{CHOOZfinal}).  
Using this as our baseline, we assume a neutron capture time 
window of $Delta t<200 \mu$s.  Based on our Monte Carlo (see
Sec.~\ref{MC}), this results in a failures to associate the neutron capture 
with the parent event $0.1$\% of the time due to late captures.  We 
address how these events can be removed in Sec.~\ref{nubarp}.

Neutron capture on Gd results in a cascade of photons with 5.6 to 8~MeV
of deposited energy, depending on the Gd isotope.  The dominant
cross sections are for $^{155}$Gd and $^{157}$Gd, both of which result
in $\sim$8~MeV of released energy.    The remaining Gd
isotopes represent $<9\!\times\!10^{-5}$ of all Gd captures.  Those
which released less than 6 MeV ($^{158}$Gd and $^{160}$Gd) represent
only $1.4\!\times\!10^{-5}$ of all Gd captures.  Therefore we assume that
the reconstructed energy for neutron captures on Gd is always $>5$~MeV.

CHOOZ found that the percentage of events which capture on Gd is 84\%~\cite{CHOOZfinal}.  We use this capture fraction in our
calculations.  To increase the probability of Gd capture, it may be
preferable to use isotopically-enhanced Gd.  This has not been done in
past experiments and requires further investigation.

The remaining 16\% of neutrons will capture on hydrogen, resulting in
a single 2.2 MeV $\gamma$.  In this analysis, it is necessary for a
large fraction of these events to be identified using a combination of
timing and position.  CHOOZ studied a trigger~\cite{CHOOZNN} for the H
capture events, but this was developed late in the experiment and not
implemented before data taking ended.  However, the initial results
looked promising.  KamLAND quotes an efficiency for $\overline{\nu} p$
events of 78$\pm$1.6\%~\cite{kamlandprl}.  This experiment is performed
on oil with no Gd-doping, thus the neutron path length is large.  The
inefficiency is largely driven by the cut on relative position of the
neutrino and neutron vertex.  In order to proceed with background
calculations, we assume a search window of 30 cm or five neutron
path lengths will be feasible.  This yields a 0.6\% inefficiency due to
neutrons which exit the window, which we will consider in
Sec.~\ref{nubarp}.  It is desirable to make this window larger, if
the trigger rate can be tolerated.

\subsubsection{Measuring the Gd and H Capture Fractions}

CHOOZ measured the capture fraction on Gd with a 1\% error using 
an Am/Be triggered neutron source.
Because this experiment will run for three times the CHOOZ
period, and because the Gd capture fraction can be measured in both near
detectors and the four far detectors (see Sec.~\ref{killbeta}), more
than an order of magnitude more calibration data will be collected.
Thus, in principle, the Gd capture fraction can be measured to better
than 0.25\%.  

As additional assurance that the capture fraction can 
be measured well, we propose a small, dedicated detector with
excellent energy resolution to accurately measure the fraction of Gd
captures.  The detector must have excellent energy resolution assuring
a clean separation between the H capture energy peak and the
Gd capture energy peak.  The detector will consist of Gd doped
scintillator which is of the same batch as the near and far detectors.
A permanently installed Am/Be source would provide the trigger.  One
would want the fiducial radius to be at least six neutron
path lengths, or 36~cm.  It need not be a miniature version of the
near detector -- in fact other designs may be preferable and easy to
obtain.  For example, the SciBath detector design proposed by the
FINeSSE experiment could be used for this purpose~\cite{FINeSSE}.

\subsubsection{\label{contam_intro}Contamination in the Detector}

As discussed in Sec.~\ref{introdesign}, the main decay chain of
concern is $^{232}$Th.  We will show in Sec.~\ref{environ} that the
fiducial volume cut reduces the background from the tank walls,
phototubes, and acrylic vessel to a negligible level.  Nevertheless,
precautions at the level of KamLAND should be taken with these
components.

The most important contamination issue is the amount of Th dissolved in 
the oil.  A small fraction of the daughters in the $^{238}$U decay chain 
also produce visible energy in the 3 to 5~MeV region.  Other radioactive 
contaminants, such as $^{40}$K and $^{14}$C are not considered because the visible energy from these decays is below the energy level of this study.  
The far detectors will be used to study contaminants, so all detectors
will be filled with oil from the same batch to assure consistent
purity.

Our goal is to achieve the same fractional Th concentration in the 
scintillator as has been achieved at KamLAND~\cite{Nakamura:2004cb}, which is 
$5.2\!\times\!10^{-17}$~g/g.  While we will show in Sec.~\ref{baddope} that 
two orders of magnitude higher contamination can be tolerated if 
necessary, KamLAND level purity is undoubtedly desirable.  Reaching 
this level of purity requires addressing the cleanliness of the 
scintillator oil and also the contamination of the Gd-dopant.  Purifying scintillator to attain low levels of dissolved thorium has been 
demonstrated.  
On the other hand, additional study is needed to assure the required 
purity of the Gd, which is isolated from contaminants by an evaporation 
process.  For this discussion we assume that $5\!\times\!10^{-17}$ g/g of 
$^{232}$Th can be attained in the detector, although we will show that 100 
times this rate can be tolerated.

Our goal for $^{238}$U contamination is the KamLAND level of
$3.5\!\times\!10^{-18}$ g/g~\cite{Nakamura:2004cb}.  
The issue of contamination of the Gd must be addressed to 
achieve these goals.

\subsubsection{\label{muonid}Cosmic Ray Identification Systems}

As described in Sec.~\ref{introdesign}, cosmic ray background must
be reduced for this analysis.  For our assumed overburden we calculate 
a cosmic ray flux, of 0.5~$\mu$/m$^2$/s or 3.5~Hz/detector.  

For the oscillation experiment, most designs propose an active veto
region which surrounds the detector~\cite{link}.  This can be
designed with at least $>99.99$\% efficiency as achieved in MiniBooNE~\cite{MBrunplan}.  

For the purpose of identifying cosmic rays in the tank, we assume that
200 photoelectrons (PE) are detected per MeV of energy deposited in the detector.  This rate of PE/MeV is similar to the CHOOZ 
design~\cite{CHOOZfinal} and is one third less than the Borexino test 
detector, CTF~\cite{CTFres2}.  It is more than sufficient for our needs here.

Cosmic rays in the detector are the unique source of events above the 
Michel electron cutoff (about 52 MeV), and are, therefore, easily 
identifiable.  Cosmic ray muons deposit about 2~MeV/cm~\cite{MBrunplan, Groom}.  This yields 0.4 PE/cm/phototube for muons.  For this analysis, we are 
interested in muons which penetrate into the fiducial volume.  To penetrate 
into the fiducial region, the muons must pass through a minimum of 70 cm of
scintillator, depositing 140 MeV of energy in this model, well above
the Michel electron cutoff.  We will call a muon with $E>140$ MeV a
``penetrating $\mu$'' for the remainder of the discussion.

The need to simultaneously reconstruct reactor-induced events and
penetrating $\mu$'s implies that the electronics must be sufficient to
reconstruct events which range from 1~MeV to at least 140~MeV.
Ability to resolve energies above 140~MeV is desirable, since it will
allow better understanding of the cosmic rays.  With 200~PE/MeV and
1000 phototubes, 140~MeV represents 28~PE/phototube.  Thus, the electronics 
requires a minimum dynamic range of at least a factor of $\sim$30 
({\it i.e.} from 1~PE to more than 28~PE) without saturation. 
The electronics used in SNO~\cite{SNO} had a dynamic range of 1 
to 1000~PE, so a substantially larger range is certainly possible.

The absence of a hit in the lower portion of the veto can be used to 
select events where the muon stopped in the detector.  This will reduce 
the rate of potential parent cosmic rays for the $^{12}$B search to an 
acceptable level and also allow this source to be isolated for calibration.  This veto system should be segmented, so that when used in conjunction 
with the upper veto, the cosmic ray track direction can be reconstructed 
to within a few centimeters.  This also reduces false coincidence rates.

\subsection{\label{cuts}The Standard Cuts}

\begin{sidewaystable}
\begin{center}
\begin{tabular}{|c|c|c|} \hline
{\bf {\normalsize Cuts, all samples}} & {\bf {\normalsize Range Retained}} & {\bf \bf \normalsize Primary Motivation }\\ \hline \hline
Fiducial volume  & 45 cm inward from Gd boundary  & maintain high $n$ efficiency.\\ \hline 
\hline 
{\bf \normalsize Vetoes, all samples} & {\bf \normalsize Description}  &{\bf \normalsize Primary Motivation} \\ \hline
Stopping Muon Veto & No lower veto hit \& penetrating $\mu$; & vetoes $^{12}$B \& Michel $e^-$ \\
                   & veto window: 260 ms &  (deadtime: 1.9\%)  \\ \hline
Muon-Hadron Veto   & Through-going $\mu$ with more energy than expected  & vetoes isotopes which $\beta$-decay \\ 
                   & from ionization or a neutron capture within 600 ms; &  (deadtime: 4.2\%) \\
                   & veto window: 3 s & \\ \hline \hline 
{\bf \normalsize Cuts, elastic scattering sample} & {\bf \normalsize Range Retained} & {\bf \normalsize Primary Motivation} \\ \hline
Minimum visible energy & $>3$ MeV & Reduce all sources \\
Maximum visible energy & $<5$ MeV & of backgrounds \\  \hline
Neutron capture energy & $< 1.8 MeV$ & identify and cut $\overline{\nu} p$ \\  
~~~~ \&  delay window & $\Delta T<200\mu$s &  \\ \hline \hline 
{\bf \normalsize Cuts, normalization sample} & {\bf \normalsize Range Retained} & {\bf \normalsize
Primary Motivation} \\ \hline
Neutron capture energy & $E>5$ MeV & Isolate well-identified events to \\ 
~~~~ \& delay window & $\Delta T<200\mu$s & maximize $n$ id purity \\ \hline
 Minimum visible energy & $>2.2$ MeV & isolate events with flux which  \\
                        &          & overlaps $\overline{\nu} e$ signal \\ \hline
\end{tabular}
\caption{\label{tab:cuts}General motivation for the major cuts.  ``Elastic'' and ``Normalization'' samples are described in Sec.~\ref{introdesign}.
Visible energy refers to the measured energy of the primary interaction.
} 
\end{center}
\end{sidewaystable}

Based on the detector described above, we propose a set of
analysis cuts.  These will be used to evaluate the
capability of the experiment.  
The cuts fall into four categories:
cuts applied to both event samples; vetoes applied to all event
samples; cuts applied to isolate the elastic scattering sample; and
cuts applied to isolate the normalization sample.  The standard cuts
on which we will base our estimate for the error on $\sin^2 \theta_W$
are listed in Table~\ref{tab:cuts}.

\subsubsection{\label{vetoes}Overview of Analysis Level Vetoes}

Two analysis-level vetoes are employed: the stopped muon veto and the
muon-hadron veto.  The first veto is designed to reduce background
from Michel electrons and $^{12}$B beta-decays.  The second removes
high-energy-muon-induced $\beta$-decaying isotopes.  

The stopped muon veto is applied in the following way.  The presence
of a ``stopped muon'' is identified by requiring a penetrating $\mu$
in coincidence with no exiting signal in the lower veto system.  All
subsequent events in a 260~ms window are then eliminated.  The 260 ms
window was chosen because it is about twelve times the $^{12}$B half-life and
thousands of muon lifetimes.  It thereby effectively eliminates
the stopped $\mu$ backgrounds.

For this discussion, we will assume the stopped muon veto is 100\%
efficient.  Inefficiency in identifying the muon signal could cause
this veto to fail.  However, this is expected to be negligible.  
Inefficiency in the lower veto increases the
deadtime, as discussed below, but does not cause the veto to fail.
Noise in the lower veto in coincidence with a stopping cosmic could
cause the veto to fail.  But a combination of selecting quiet
phototubes and segmented construction can reduce this to a negligible
level.

The muon-hadron veto removes all events in a 3 second window following 
a though-going muon accompanied by either a significant energy 
deposition over that expected from ionization alone, or at least one 
captured neutron within 600~ms.  The purpose is to reduce the 
background from muon-induced $\beta$-decaying isotopes.  The production 
of these isotopes is typically accompanied by a sizable hadronic shower 
and an average of 3 free neutrons~\cite{Wang} (in many cases the 
neutrons are fragments of the parent $^{12}$C isotope, see Table 
\ref{isotopesource}).  For the purposes of this study we assume that the 
muon-hadron veto will be 95\% efficient.

\subsubsection{Deadtime Induced by the Vetoes}

Deadtime is not a major consideration in this analysis because the
signal and normalization samples will be equally affected.  Nevertheless, 
when one is performing a precision measurement, as a
matter of practice it is best to have the lowest possible deadtime.
Also, one aims for a small deadtime so that one can run for the
minimum possible time.  In considering the discussion below, note that
a veto which depends only on the presence of a cosmic ray, without
asking for a stopping signal or accompanying hadronic activity, would lead to
intolerable deadtimes.

The deadtime for the stopped muon veto will be 1.1\% given the
expected stopped muon rate of 0.042 Hz.  This is acceptable.  In
principle, inefficiency in the lower veto could produce
misidentified stopping muons.  It is reasonable to assume that
this veto can be made better than 99\% efficient.  Assuming a 1\%
inefficiency would lead to a deadtime of only 1.9\% (veto inefficiency
and real stopped muon rate, combined).  This is sufficiently small that
it is not an issue.

To calculate the rate at which the muon-hadron veto will fire, one
needs to consider both the muon rate and the neutron capture
rate.  To estimate the neutron rate, we use a simulation which is
described in Sec.~\ref{MC}.  We expect 0.042 muon-induced
neutrons/s.  However, we note that many of these neutrons will be 
produced in association with the same cosmic muon.  To correct for 
multiple neutron production, we use the calculated average 
multiplicity of three.  Thus we take as our prediction a rate of 0.014 Hz.
Opening a 3~s window, thereby introduces a 4.2\% deadtime.  This is an
acceptable rate and is probably an overestimate.  In fact, to further
reduce backgrounds from $\beta$-decaying isotopes, one might consider
enlarging this window.

To address the rate of accidental firing of this veto, we must
consider the types of events which can cause each component to fire.
The cosmic signal is unique among the types of events which can occur,
due to the very high energy.  Therefore, we assume that there is no
accidental background in this component.  The most likely
false vetoes come from Michel electrons and $^{12}$B decays, because
these events are correlated with an incoming muon.  Neutrons which enter
the tank can also produce a muon-neutron coincidence.  This rate is
much lower, however, because the neutron and cosmic ray are not
correlated, and so we do not consider this here.

\subsection{\label{MC}Calculation of Neutron Production and Transport}

The above discussion and that which follows relies on a simulation of 
the interactions of cosmic rays muons at the expected overburden.  To 
calculate the production of fast neutrons  we begin with a 
parameterization of the muon rate at the surface as a function of energy 
and zenith angle~\cite{PDG}.  The muon rate is divided into 750,000 bins 
in energy (100 MeV steps from 0 to 2.5 TeV) and angle ($2^{\circ}$ steps
from $0^{\circ}$ to $60^{\circ}$).  In each bin the energy is
attenuated over steps of one meter (larger steps are used for very high
energies) according to the average energy loss as a function of muon
energy~\cite{Groom}.  The muon rate and spectrum at the given depth are used to
determine the neutron spectrum and rate following the neutron production model of Wang {\em et al.}~\cite{Wang} .  Similarly, the isotope
production is determined using the normalization and energy scaling of
Hagner {\it et al.}~\cite{Hagner}.  The production rates for $^9$Li, $^8$He, $^8$Li, $^6$He, $^9$C, and $^8$B come from the measured rate of Hagner {\it et al.}, and the $^{12}$B rate comes from an observation made by KamLAND~\cite{Araki:2004mb}.

The neutron transport Monte Carlo takes the neutron production energy
distribution as an input, and propagates the neutrons assuming elastic
scattering in $CH_2$ with 0.1\% Gd by weight.  The cross sections for
elastic scattering and capture on H, C, and Gd are taken from 
reference~\cite{BarnBook}.  These calculations are then used in determining the efficiency of the muon neutron veto and inverse beta decay rejection.

\section{\label{nubare}$\overline{\nu} e$ Event Rate and Identification}

Neutrino-electron scattering measurements have been studied for many 
years~\cite{Imlay}. $\overline{\nu} e$ events result either from scattering 
via exchange of a $Z$ boson, or annihilation via exchange of a $W$ boson. 
The differential cross section for $\overline{\nu}_{e}e^{-}$ scattering
is:
\begin{eqnarray*}
\frac{d\sigma}{dT}\!&\!=\!&\! 
\frac{\pi\alpha^{2}\mu_{e}^{2}}{m_{e}^{2}}\frac{\left(1-T/E_{\nu}\right)}{T} +
\frac{G_{F}^{2}m_{e}}{2\pi} \times \\
& &\!\left[ \left(  g_{V}\!+\!g_{A}\right)^{2}\!+\!\left( g_{V}\!-\!g_{A}\right)  ^{2}\!\left(\! 1\!-\!\frac{T}{E_{\nu
}}\!\right)^{2}\! +\!\left(  g_{A}^{2}\!-\!g_{V}^{2}\right)\!  \frac{m_{e}T}{E_{\nu}^{2}}\!\right] 
\end{eqnarray*}
where $E_{\nu}$ is the incident $\overline{\nu}_{e}$ energy, T the electron recoil kinetic energy, and the couplings are given by~\cite{Vogel:1989iv}
\[
g_{V}=2\sin^{2}\theta_{W}+\frac{1}{2}\text{~~~~ }g_{A}=-\frac{1}{2}.
\]
The term in brackets is the weak interaction contribution, and the
last term gives the contribution from electromagnetic scattering if
the neutrino has a magnetic moment, $\mu_{e}$.

The total visible energy, $E_{vis}$ in elastic scatters is the kinetic
energy of the $e^-$, $T$. This is in contrast to $\overline{\nu} p$ events
where additional visible energy will come from both the positron
annihilation and the neutron capture.

If one could reconstruct both $T$ and $E_\nu$ in $\overline{\nu} e$ events, 
then an analysis of the $T/E_\nu$ dependence would be attractive.  This 
method evades the issue of absolute normalization.   However, in the 
generic detector described above the angle of the $e^-$ cannot 
be reconstructed.   Therefore, only $T$ is measurable.   Once this 
cross section is folded with the reactor flux, the variation of 
the shape versus $T$ is insensitive to $\sin^2 \theta_W$.   

On the other hand, the total rate of $\overline{\nu}$ events is sensitive to
$\sin^2 \theta_W$.  In fact, the sensitivity to $\sin^2 \theta_W$ can
be enhanced by introducing a cut on $T$.  Integrating over the recoil
electron kinetic energy from $T_{\min}$ to
$T_{\max}$ gives a cross section as a function of $E_{\nu}$ given by:

\begin{eqnarray*}
\sigma\mid_{T_{\min}}^{T_{\max}}  &  =\frac{G_{F}^{2}m_{e}}{2\pi}\!\left[\!
\left(  \left( g_{V}\!+\!g_{A}\right)  ^{2}+\left( g_{V}\!-\!g_{A}\right)
^{2}\right)  \left(  T_{\max}\!-\!T_{\min}\right)  \right. \\
&  +\frac{1}{2}\!\left(  \frac{m_{e}\left(  g_{A}^{2}\!-\!g_{V}^{2}\right)  }%
{E_{\nu}^{2}}\!-\!\frac{2\left(  g_{V}\!-\!g_{A}\right)  ^{2}}{E_{\nu}}\right)
\!\left(  T_{\max}^{2}\!-\!T_{\min}^{2}\right) \\
&  \left. \!+\!\frac{\left(  g_{V}\!-\!g_{A}\right)  ^{2}}{3E_{\nu}^{2}}\left(
T_{\max}^{3}\!-\!T_{\min}^{3}\right)  \right]
\end{eqnarray*}
$T$ in the range of 2.5 to 5 MeV is optimal.  As discussed in
Sec.~\ref{introdesign}, however, cuts on various backgrounds dictate
$3<(T=E_{vis})<5$ MeV.

We assume that the term associated with the neutrino magnetic moment
($\mu_e$) is negligible, based on astrophysical constraints~\cite{Fukugita:1987uy, Lattimer:1988mf, Barbieri:1988nh, Raffelt:gv}.
We note, though, that the lab-based limits on the
neutrino magnetic moment are two orders of magnitude higher~\cite{munu}.  
If the neutrino magnetic moment were
just below the lab-based limit ({e.g.  $10^{-10}$}), then this term
would result in a 12\% increase in the elastic scattering rate.

\section{\label{nubarp}The $\overline{\nu} p$ Background}

A major potential source of background comes from misidentified
$\overline{\nu}_{e}p\rightarrow e^{+}n$ events.  The
cross section is given by~\cite{Vogel:1999zy}:
\[
\sigma\left(  E_{e^{+}}\right)  =\frac{2\pi^{2}}{m_{e}^{5}f\tau_{n}}p_{e^{+}%
}E_{e^{+}}%
\]
where $E_{e^{+}}(p_{e^{+}})$ is the energy(momentum) of the outgoing
positron, $f=1.71465(15)$~\cite{GrattaRMP, CHOOZfinal} is the free
neutron decay phase-space factor, and $\tau_{n}=886.7\pm1.9$ s~\cite{nlife} is the neutron lifetime.  Using these measurements of $f$ and $\tau_n$, one 
finds that the cross section is known to $\sim$0.2\%.  For this process,
the $E_{\overline{\nu}_{e}}$ energy threshold is 1.804 MeV and the
incoming $\overline{\nu}_{e}$ energy is simply related to the outgoing
positron energy by
\[
E_{\overline{\nu}}=E_{e^{+}}+(M_{n}-M_{p})=E_{e^{+}}+1.2933\text{ MeV}~.%
\]

Most $\overline{\nu} p \rightarrow e^+ n$ are identified by the outgoing
neutron.  However, the neutron may not be observed, because of the the 
inefficiency on triggering on H capture, or because the neutron was
outside of the neutron-delay time window.  In the latter case, $0.1$\%
of the neutrons will capture late, leaving a primary neutrino vertex and
a secondary neutron vertex which is mistakenly unassociated.
However, the $E_{vis}$ window requirement further reduced the fraction of these events that contribute to as background to the $\overline{\nu} e$ sample.  

\subsection{\label{nreject}Rejection through $n$ Identification}

Most $\overline{\nu} p$ events can be rejected through identification of the
time-delayed $n$.  We take the efficiency for reconstructing the
photons from Gd capture to be 100\%.  On the other hand, the
efficiency for identifying the photon associated with H capture is
only $99.4$\%.  Thus the total $n$-identification efficiency is $0.84
+ (0.16\times 0.994) = 0.999 $.  This is to say, $0.1$\% of $n$ events
within the neutron time window will fail to be identified.

The systematic error on this is small.  The error on the Gd capture
fraction is assumed to be 0.3\% (see Sec.~\ref{generic}).  The Gd
capture fraction introduces an error, $\delta$, which changes the
efficiency to $\epsilon=(0.84 + \delta)+(0.16-\delta)\times 0.994$.
Therefore $d\epsilon = 0.006\times d\delta \simeq 2\!\times\!10^{-5}$.  
From binomial statistics, one needs only 17,000 tagged calibration-source 
events to obtain a 10\% error on this inefficiency.  This should be
achievable using the {\it in situ} Am/Be source calibration. We therefore
assume no systematic error contribution from this source.

\subsection{\label{Ereject}Rejection through an $E_{vis}$ Cut}

There are three cases where the neutron is ``lost'' to the analysis.
First, the neutron capture occurs outside the neutron delay time
window in 0.1\% of the cases.  Second, $0.0011$\% of the time the neutron 
exit the Gd-doped central region without capturing.  Third, for $0.1$\% of 
the time the neutron captures on hydrogen, but it is outside the 
H-trigger spatially allowed range (see Sec.~\ref{reactorevents}).  The 
rates for these potential sources of background are all
further reduced by the $E_{vis}$ cut.

In each of these cases only the positron energy is observed.  
Approximately 45\% of the positron events fall within the $E_{vis}$~MeV 
window.  The background for late captures is, therefore, 0.045\% of all 
$\overline{\nu} p$ interactions.  The fraction of events with a neutron 
which exit and with neutrino vertex energy in the window is $7\!\times\!10^{-6}$.  The case where the neutron is lost from the 
H-trigger search region, follows the same argument.  This background 
source is, therefore, 0.045\% of the $\overline{\nu} p$ events.  Again we 
assume no appreciable systematic error on these values.

Based on this, we estimate the the total $\overline{\nu} p$ background to be
0.09\% of the $\overline{\nu} p$ interactions with a negligible systematic 
error.  

\section{\label{environ}Environmental Backgrounds}

Section~\ref{introdesign} introduced the environmental backgrounds
which reactor experiments face.  Potentially, they come from
contaminants, cosmic ray muons, and products of cosmic ray muons such as
Michel electrons, spallation neutrons, and muon-produced isotopes.  We
assume cosmic muons are readily identifiable, as described in
Sec.~\ref{muonid}.  We show that the $3<E_{vis}<5$~MeV window, in
combination with the vetoes proposed in Sec.~\ref{vetoes}, reduce
most environmental backgrounds to a negligible level.  

\subsection{Sources}

\subsubsection{$^{238}$U and $^{232}$Th decay chains} 

\begin{figure}
\vspace{5mm}
\centering
\includegraphics[width=8.5cm]{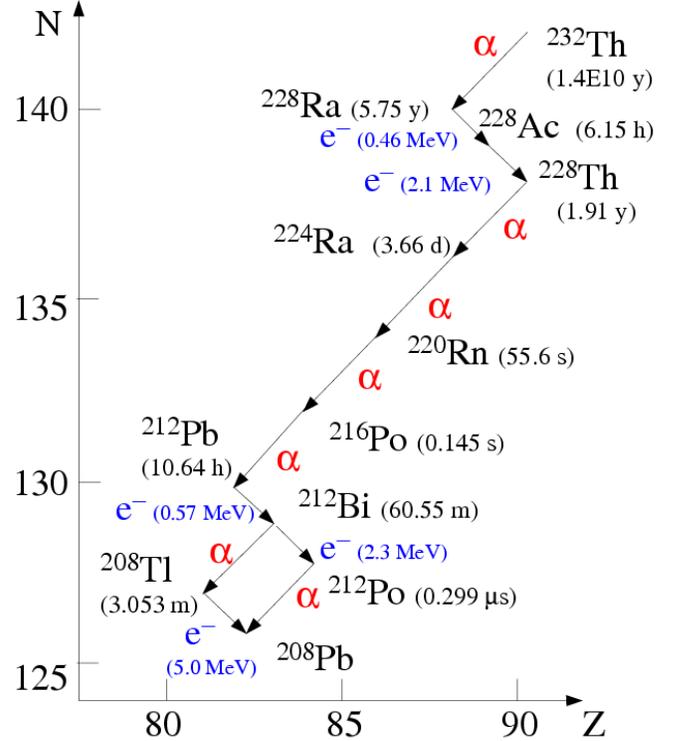}
\vspace{1mm}
\caption{\label{thoriumchain}Decay chain for $^{238}$Th.  The half-life of 
each step is listed in parentheses.   For $\beta$ decays, the total visible 
energy released ($\beta$'s and $\gamma$'s) is noted.}
\end{figure}
 
The decay chain for $^{232}$Th is shown in Fig.~\ref{thoriumchain}.
The half-life for each step is also listed in the figure.   

The $^{232}$Th chain produces six alpha particles of energy 4 to 9 MeV.
Scintillation signals quench by a factor of 10 to 15, so each alpha
deposits roughly 0.25 to 0.8 MeV in the detector.  It is highly
unlikely that multiple decays will occur simultaneously, thus the
$\alpha$'s do not represent a background.

Five $\beta$'s are also produced in the decay chain, and the
respective energies of the $\beta$'s are listed on
Fig.~\ref{thoriumchain}.  The issue for this analysis is the
decay of $^{208}$Tl to $^{208}$Pb.  This releases a $\beta$ with
energy up to 1.8 MeV and, simultaneously, $\gamma$'s, including a 
2.4~MeV $\gamma$.  The total energy of this decay is 5 MeV.  Therefore
this decay has sufficient energy to appear in the 
$3<E_{vis}<5$~MeV window.   This decay lies on a branch of
the chain, such that only 35\% of the parent $^{232}$Th will result this
decay.

The Th-related contaminants on the acrylic vessel will not result in
background because the events will be removed by the fiducial volume
requirement for this analysis.  Also, any $\gamma$'s produced by this decay
chain, which may enter the fiducial volume, will be below the 3~MeV
visible energy window, which means that they will not contribute.  
Therefore, we only consider the $^{232}$Th dissolved 
in the scintillator.

Calculating from our assumed $^{232}$Th concentration of 
$5\!\times\!10^{-17}$ g/g, and half-life of  $1.4\!\times\!10^{10}$~years 
we expect to have 183 decays of each isotope in the chain between the two 
detectors.  But only 35\% of the chains go through the $^{208}$Tl branch, 
and only 51\% of those will have a visible energy on the 3 to 5~MeV window.  Therefore, we expect only 93 $^{232}$Th background events in all.

As seen in Fig.~\ref{fig:UTh}, there is also a small fraction of
background events, from 3 to 3.5 MeV, which are due to the $^{238}$U
chain.  These events are produced from a $\beta$ decay within this chain
with an endpoint of 3.2 MeV.  We will assume that we can achieve the
same level of uranium contamination as KamLAND.  
Judging from Fig.~\ref{fig:UTh}, the U contribution in the
visible energy window is only about 5\% of the Th contribution,
or about 5 events for the two near detectors combined.

The systematic error on these backgrounds can be determined by two
methods. In the first method, samples of the oil will be
studied in a low background counting facility located deep
underground.  This will allow a precise absolute measure of the $\beta$ 
decay of concern.  However, it relies on Monte Carlo to accurately 
represent the smearing.  In the second method, the far detectors are used 
to measure the background expected in the near detector.  This method is
discussed in Sec.~\ref{killbeta}.

\subsubsection{\label{michel}Michel Electrons}

The ratio of stopping to through-going muons at 300 mwe has been
calculated to be $6\!\times\!10^{-3}$/m~\cite{Cassiday}.  The flux of 
cosmic rays entering a detector under 300 mwe overburden 
is 0.5 $\mu$/m$^2$/s.  The stopping rate in the detector is therefore
0.042 Hz, or about $3.3\!\times\!10^6$ Michel decays per detector for 
the run.

The stopped muon veto represents many thousands of muon lifetimes, and
so the background from Michel electrons is negligible in this
analysis.  Instead,  Michel electrons represent a well-identified
control sample for studies in this analysis.  If all detectors are
built identically, then the Michel samples can be combined, greatly
enhancing these studies.

\subsubsection{\label{inneut}Neutrons}

Cosmic rays can produce spallation neutrons either outside or within
the tank, but these are a negligible background. This is because
these neutrons, like unassociated neutrons described above, will fail
the energy window.  H~captures result in a visible energy which is 
below 3~MeV, and Gd captures result in a visible energy above 5~MeV\@.  
Moreover, for neutrons which enter the tank, the $n$ must traverse 6.7 
interaction lengths (40 cm) of Gd-doped scintillator in order to reach 
the fiducial volume.  The resulting rejection is about 0.001.  We 
therefore take this background to be negligible in this study.

\subsubsection{\label{boron}Stopping-muon-induced $^{12}$B}

Cosmic ray muons entering the detector may capture and produce
$^{12}$B which $\beta$ decays.  Capture occurs 8\% of the time in oil~\cite{Suzuki}.  Because only the $\mu^-$ captures, we gain a factor of
two.  In addition, only 19\% of these events, appear
in the $E_{vis}$ window, and only about 17\% of $\mu^-$ captures on 
carbon result in the ground state of $^{12}$B~\cite{Suzuki,Miller1972}.  
Therefore, the rate of $^{12}$B decays from $\mu^-$ capture is about 
$5\!\times\! 10^{-5}$~Hz.  This represents about 4,000 events per detector 
during the run, which is too high a level of background for this analysis.  
To reduce the rate further, we introduced the stopping muon veto, which was 
described in Sec.~\ref{vetoes}.  The 260~ms veto window is about twelve 
times the half-life of $^{12}$B.  We expect less than one $^{12}$B 
background events from $\mu^-$ capture between the two near detectors.
 
\subsubsection{\label{splat}High-energy-muon-induced Isotopes}

\begin{table}[tbp]
\begin{center}
\begin{tabular}{|c|c|c|c|c|} \hline
Isotope  &  Half-life  & Endpoint & Rate   & Rate         \\
         & (s)        & (MeV)    & (/t/d) & (detector/d) \\ \hline
$^{9}$Li+$^{8}$He &  0.18 \& 0.12 & 13.6 \& 10.6 & 0.15$\pm$0.02 & 1.95$\pm$0.26 \\
$^{8}$Li & 0.84   & 16.0  &   0.28$\pm$0.11 & 3.64$\pm$1.43 \\
$^{6}$He & 0.81   & 3.5   &   1.1$\pm$0.2   & 14.70$\pm$2.60 \\ 
$^{12}$B & 0.02   & 13.4  &   4             & 50        \\ \hline
$^{9}$C  & 0.13   & 16.0  &   0.34$\pm$0.11 & 4.42$\pm$1.43 \\
$^{8}$B  & 0.77   & 13.7  &   0.50$\pm$0.12 & 6.50$\pm$1.56 \\ \hline
\end{tabular}
\caption{\label{isotopes1}Isotopes with energy endpoint $>3$ MeV. 
Rate/detector/day is for the generic 13 ton detector under 300 mwe overburden. Top table: $\beta^-$ production; bottom table: $\beta^+$ production.}
\end{center}
\end{table}

High energy cosmic rays produce $\beta$ decaying isotopes in a number
of ways.  Spallation refers specifically to nuclear disintegration due
to interaction with a virtual photon, although the term is often used
loosely.  Other sources are elastic and inelastic scattering.  High
energy secondary neutrons and pions can also produce isotopes, so that
modeling the transport and interaction of secondaries is important.
Calculations using measured isotope
production rates by muons are in fairly good agreement with the
observations at KamLAND and can thus be used to estimate the
background rates for this measurement.  The sources for muon-induced
$\beta$-decaying isotopes of particular concern for this analysis are
listed in Table~\ref{isotopesource}.

The raw rates for a 13 ton fiducial volume scintillator-oil-based
detector located under a 300 mwe overburden are shown in column 5 of
Table~\ref{isotopes1}.  

The signal from $^{9}$Li and $^{8}$He were indistinguishable in the
NA54 data~\cite{Hagner}, and hence they are grouped together here.  
For the sake of argument here, we assume that the contributions of 
$^{9}$Li and $^{8}$He are equal for the measured rate in NA54, although 
we consider this further below.  It should be noted that if the relative 
rates of the two processes are not determined, then one needs to include 
a systematic which covers the range from the assumption of 100\% Li to 
100\% He.

\begin{table}[tbp]
\begin{center}
\begin{tabular}{|c|c|c|c|} \hline
Isotope  & $E_{vis}$ Cut & Correlated $n$ & Veto/Final Rate \\ \hline
$^{9}$Li & 0.18          & 0.09           & 0.0045          \\
         &  (19\%)       & (50\%)         & (5\%)           \\
$^{8}$He & 0.29          & 0.24           & 0.012           \\
         &  (30\%)       & (84\%)         & (5\%)           \\
$^{8}$Li & 0.47          &     N/A        & 0.024           \\  
         &  (13\%)       &                & (5\%)           \\
$^{6}$He & 0.29          &     N/A        & 0.015           \\
         &  (2\%)        &                & (5\%)           \\
$^{12}$B & 9.5           &     N/A        & 0.48            \\
         &  (19\%)       &                & (5\%)           \\
$^{9}$C  & 0.53          &     N/A        & 0.027           \\
         &  (12\%)       &                & (5\%)           \\
$^{8}$B  & 0.65          &     N/A        & 0.033           \\ 
         &  (10\%)       &                & (5\%)           \\ \hline
\end{tabular}
\caption{\label{isocuts}Isotopes decays/detector/day after each cut, applied sequentially.  Value in parenthesis is the percent of the rate retained after the cut.}
\end{center}
\end{table}

By taking the fraction of events in a 3 to 5~MeV window, assuming the
correct $\beta$-decay spectrum, we obtain the approximate rate with
the energy cut, shown in Table~\ref{isocuts}, column 2.  At the time of 
the $\beta$ decay, a neutron accompanies 50\% of the $^{9}$Li decays 
and 16\% of the $^{8}$He decays and therefore will not contribute to 
the elastic scattering background.  One therefore obtains the rates 
shown in column 3.  Finally, in column 4, we show the result of 
introducing the muon-hadron veto.   The total is 0.60$\pm$0.13 
(sys\footnote{In the absence of a reported error in  Ref.~\cite{Araki:2004mb},
the calculation of this systematic error assumes a 25\% error 
on the $^{12}$B production rate.})
events/day/detector or 1080$\pm$234 events in both detectors for the entire 
run.  This leads to a systematic error of just over 2\% in the 10000 event 
signal, which is at least a factor of two higher than can be accepted if 
the goal is to match the NuTeV errors.  Therefore, it is crucial to
constrain this systematic from the data.

\subsubsection{Neutrino Backgrounds}

Solar neutrinos also  represent a possible environmental background.
As calculated by reference~\cite{donna}, the background from solar
neutrino interactions is small.  The rate for the flux between 3 to 5
MeV, which is expected to be dominated by the $^{8}$B solar neutrinos,
is 4 events in the 900 day run.  Atmospheric neutrino rates are
considerably lower than solar neutrino rates, and so these, too, can
be neglected.  Also, geoneutrinos from $\beta$-decays in the core of 
the earth, which have energies $< 2.6$ MeV, are not an issue in this 
analysis.  

\subsection{\label{killbeta}Use of the Far Detectors to Reduce the 
Systematic Error}

The far detectors are an ideal place to cross check the predictions
for environmental backgrounds, because the signal is less than 5\% of the
background rate in the far detectors.  These backgrounds mainly come from
muon-induced isotopes, but there is also a small contribution 
of from the $^{238}$U and $^{232}$Th contaminants.  

The far detector measurements can constrain the near detector
backgrounds only to the extent that all detectors are built
identically.  The design feature which is likely to be least similar
between the near and far detectors is the overburden.  Despite the
homogeneity of the rock in, for example, the Braidwood 
area~\cite{braidwood}, the overburden for the near and far detectors may 
well differ by a few percent for shafts of identical depth.

With an overburden difference of $\sim$3\%, the rate of 
muon-induced isotopes will differ in the near and far detectors.
However, for such small variations in overburden, one expects shifts
in the normalization, with no significant deviation as a function of
energy.  We propose two methods for correcting the muon-induced 
background normalization between the near and far detectors.

The first method is to correct the normalization in the far detector
using the ratio of cosmic ray rates in the near compared to the far
detectors.  Given a cosmic ray rate of 3.5 Hz per detector, one
expects more than $3\!\times\!10^8$ cosmic ray events per year in each
detector.  The statistical error in the normalization correction
is therefore negligible.  We will assume that this is the method which is
employed, and not consider an error from the normalization correction.

The second method uses the high energy ($E> 10$~MeV) $\beta$-decays to
normalize the near-to-far detector rates.  This cut is chosen to be
sufficiently high so that in the near detector, inverse beta decay, and
elastic scattering events do not contaminate the sample.  It is 
sufficiently low, however, that we expect about 12,000 events per detector,
allowing a high statistics measurement.  It should be noted that this
sample is not contaminated by Michel electrons due to the
stopping muon veto.  Given two near detectors we expect 24,000 events.
The error on the normalization from this method is consequently about
0.6\%.  This is sufficiently small to serve as a useful cross check for
the correction based on relative cosmic ray rates.  Note that all of
the isotopes except for $^{6}$He and $^{8}$He contribute to $E>10$;
this is therefore a direct check of the dominant sources of isotopes.

The reactor-induced events in the far detector must be subtracted.  At
1.8 km from the reactor core, the signal rate and the $\overline{\nu} p$
backgrounds are both reduced by the factor $(0.2/1.8)^2\simeq0.01$.  One 
therefore expects only about 50 elastic signal events and 13 inverse 
beta decay background per detector, compared to over 540 events from 
the environmental backgrounds.

Combining the information from all four far detectors, and then 
combining this with the direct calculation of Sec.~\ref{splat},
1080$\pm$32(stat)$\pm$23(sys) events.

From all sources of environmental backgrounds (U/Th and muon-induced isotopes) we expect a total of 1178$\pm$39(stat)$\pm$26(sys) events.

\subsection{\label{overburd}Comment on Overburden}

\begin{table}[tbp]
\begin{center}
\begin{tabular}{|c|c|c|c|c|} \hline
                          &  25 mwe & 50 mwe &  300 mwe &  450 mwe \\ \hline 
muons (/m$^2$/s)  & 88.3  & 24.6   & 0.53    &  0.19 \\    
neutrons (/t/d)   &  11360 & 4942 &  322 &     145\\ \hline    
$^{9}$Li+$^{8}$He &  5.3$\pm$0.9  &   2.3$\pm$0.3 &  0.15$\pm$0.02 & (6.6$ \pm 0.7)\!\times\! 10^{-2}$ \\  
$^{8}$Li & 10.$\pm$5.   & 4.4$\pm$1.9  & 0.28$\pm$0.11 & 0.13$\pm$0.05 \\
$^{6}$He & 40.$\pm$12.  & 17.$\pm$4.5  & 1.1$\pm$0.2   & 0.50$\pm$0.07 \\
$^{12}$B & 150          & 60           & 4             & 1.8           \\
$^{9}$C  & 12.1$\pm$5.3 & 5.2$\pm$2.1  & 0.34$\pm$0.11 & 0.15$\pm$0.05 \\
$^{8}$B  & 17.9$\pm$6.4 & 7.7$\pm$2.5  & 0.50$\pm$0.12 & 0.22$\pm$0.05 \\ \hline
\end{tabular}
\caption{\label{isoburd}Isotopes with energy endpoint $>3$ MeV for various 
overburdens. Rate for the isotopes is in decays/ton/day.}
\end{center}
\end{table}

In this study we have assumed an overburden of 300 mwe equivalent.  
However, at the Braidwood site~\cite{braidwood} it is possible to reach 
450 mwe.  Larger overburden is advantageous to this analysis.  The core-to-detector distance increases by $\sim$10\%, but the rate of 
isotope production drops by more than a factor of two (see 
Table~\ref{isoburd}).  As a result, 300 mwe should be regarded as a 
minimum and the detector halls will be constructed with the maximum 
feasible overburden.

If this experiment is performed at a facility where the overburden
between near and far detectors varies substantially, so the far
detector cannot be used to constrain the near detector background rates, 
then an overburden which is deeper than 300 mwe is recommended.  One can 
use the $E>10$ MeV rates in the near detector to somewhat constrain the
errors, but this does not significantly reduce the 2\% error from the 
spallation background.  With an overburden of 450 mwe one could achieve
an error of $\sim$1\%, which is not ideal but still tolerable.

On the other hand, a shallow overburden cannot be tolerated due to the
high cosmic ray rate as shown in Table~\ref{isoburd}.  For example,
for 50 mwe, the through going cosmic ray rate increases by almost a
factor of 15.  The stopping rate will be substantially larger.  This
will lead to intolerable backgrounds from $^{12}$B and
high-energy-muon-induced isotopes.

\section{\label{norm}The $\bar{\nu} p$ Normalization Sample}

\begin{figure}
\vspace{5mm}
\centering
\includegraphics[width=7.5cm,bb=70 370 544 686, clip=true]{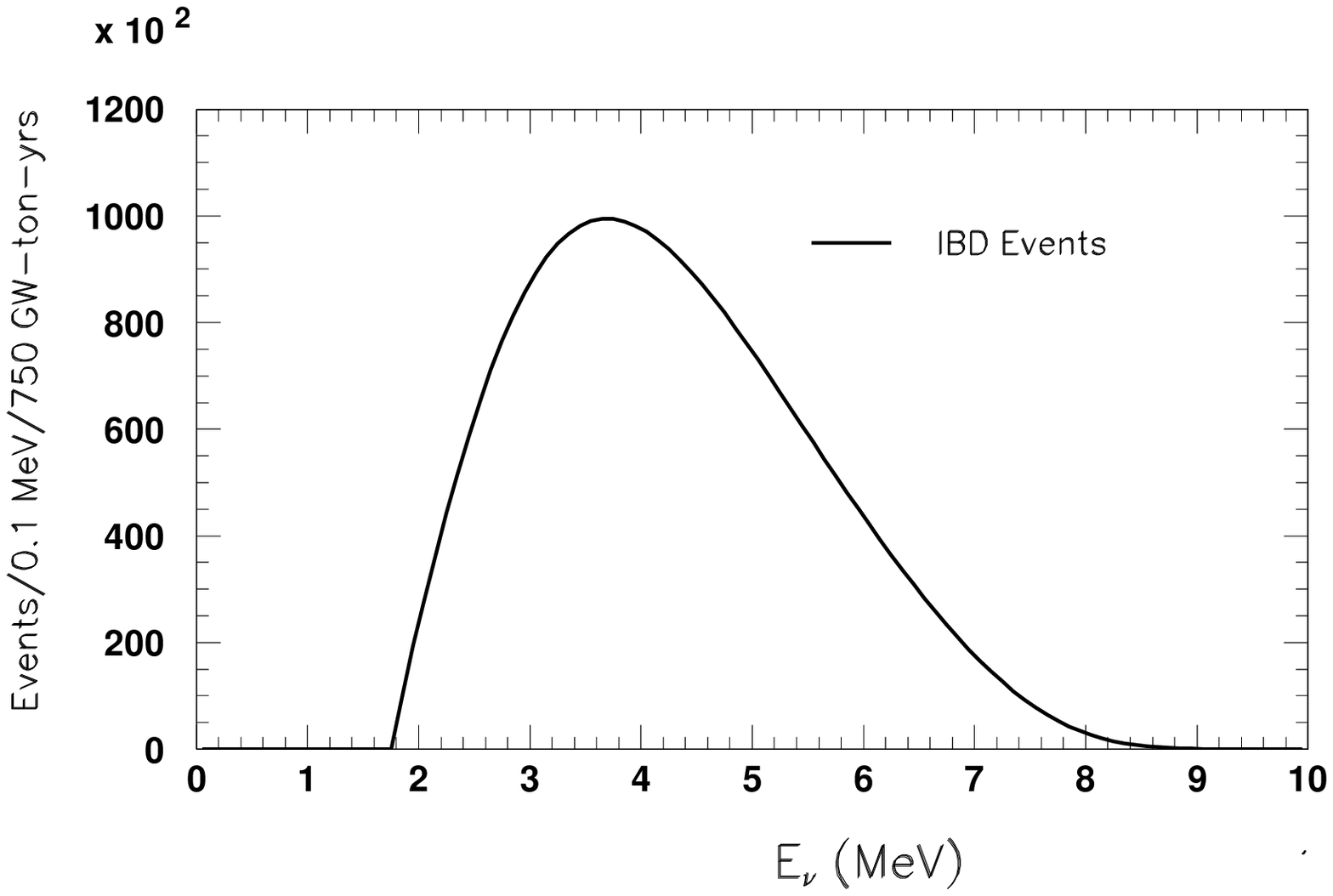}
\includegraphics[width=7.5cm,bb=70 370 544 686, clip=true]{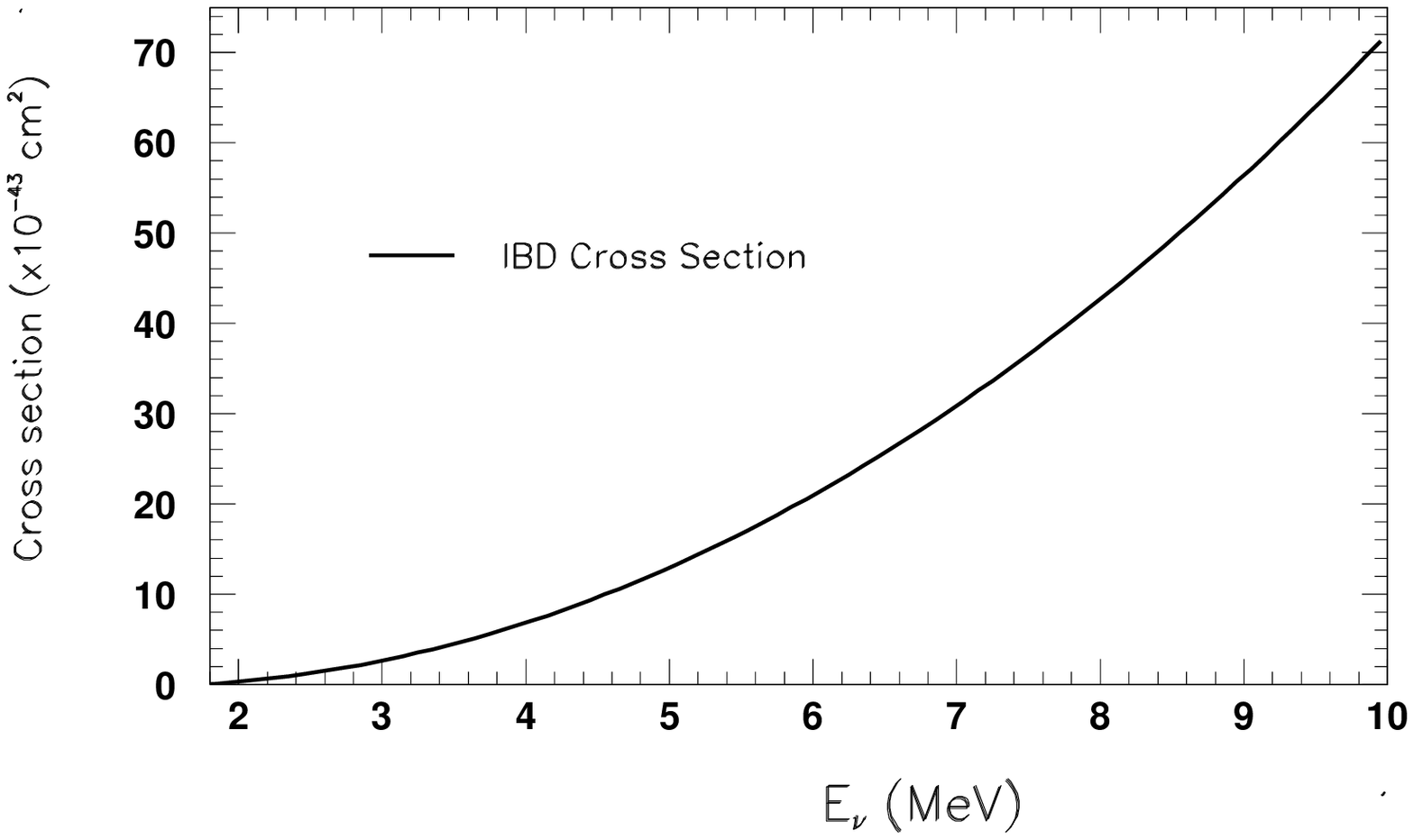}
\includegraphics[width=7.5cm,bb=70 370 544 686, clip=true]{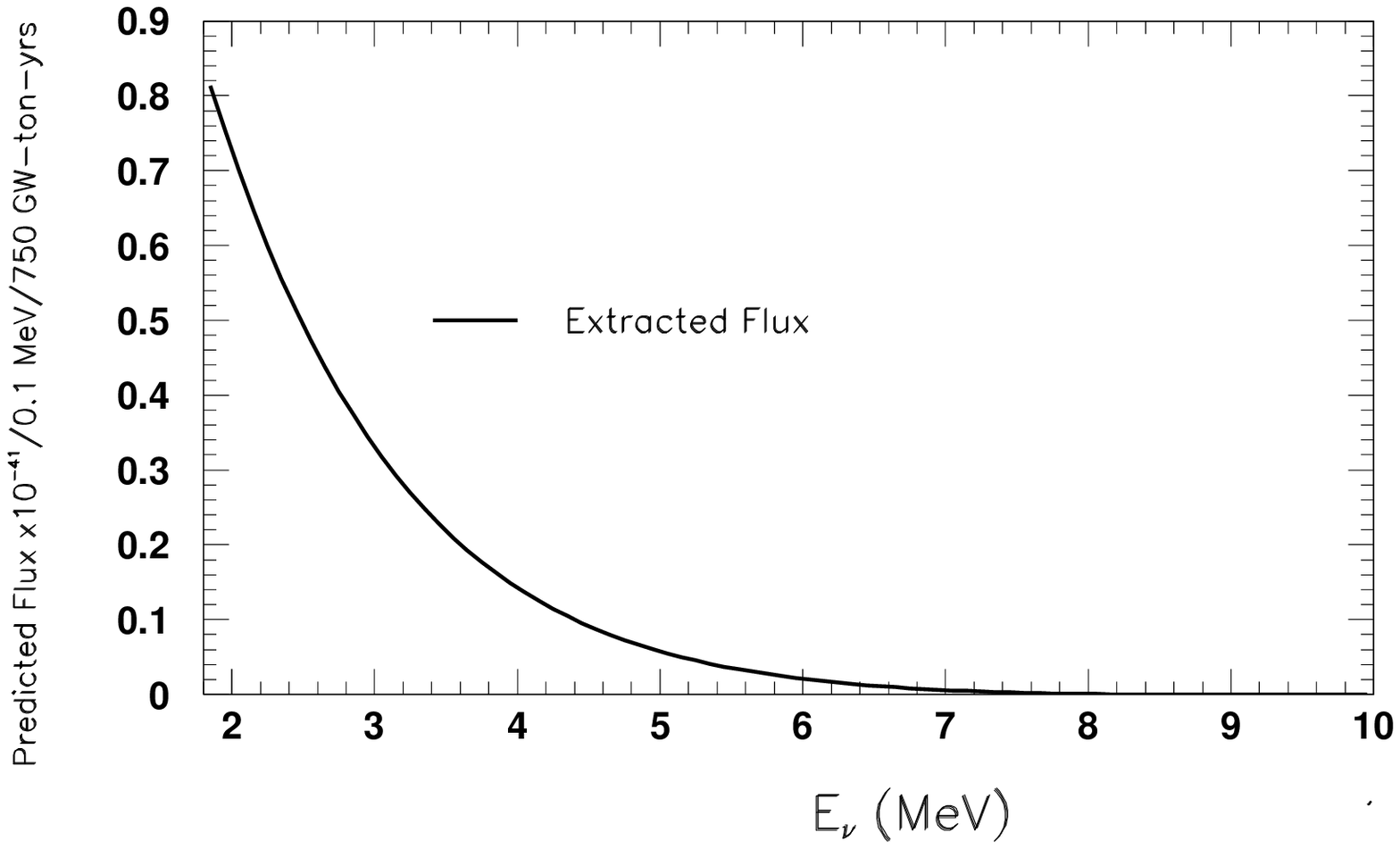}
\vspace{1mm}
\caption{\label{getibdflux}Illustration of method for extracting the $\overline {\nu}_{e}$ flux.  Top: inverse beta decay events {\it vs.} neutrino energy.  Middle:  predicted cross section for inverse beta decay events. Bottom: extracted flux obtained from dividing the event distribution by the predicted cross section.}
\end{figure}

\begin{figure}
\vspace{5mm}
\centering
\includegraphics[width=7.5cm,bb=70 370 544 686, clip=true]{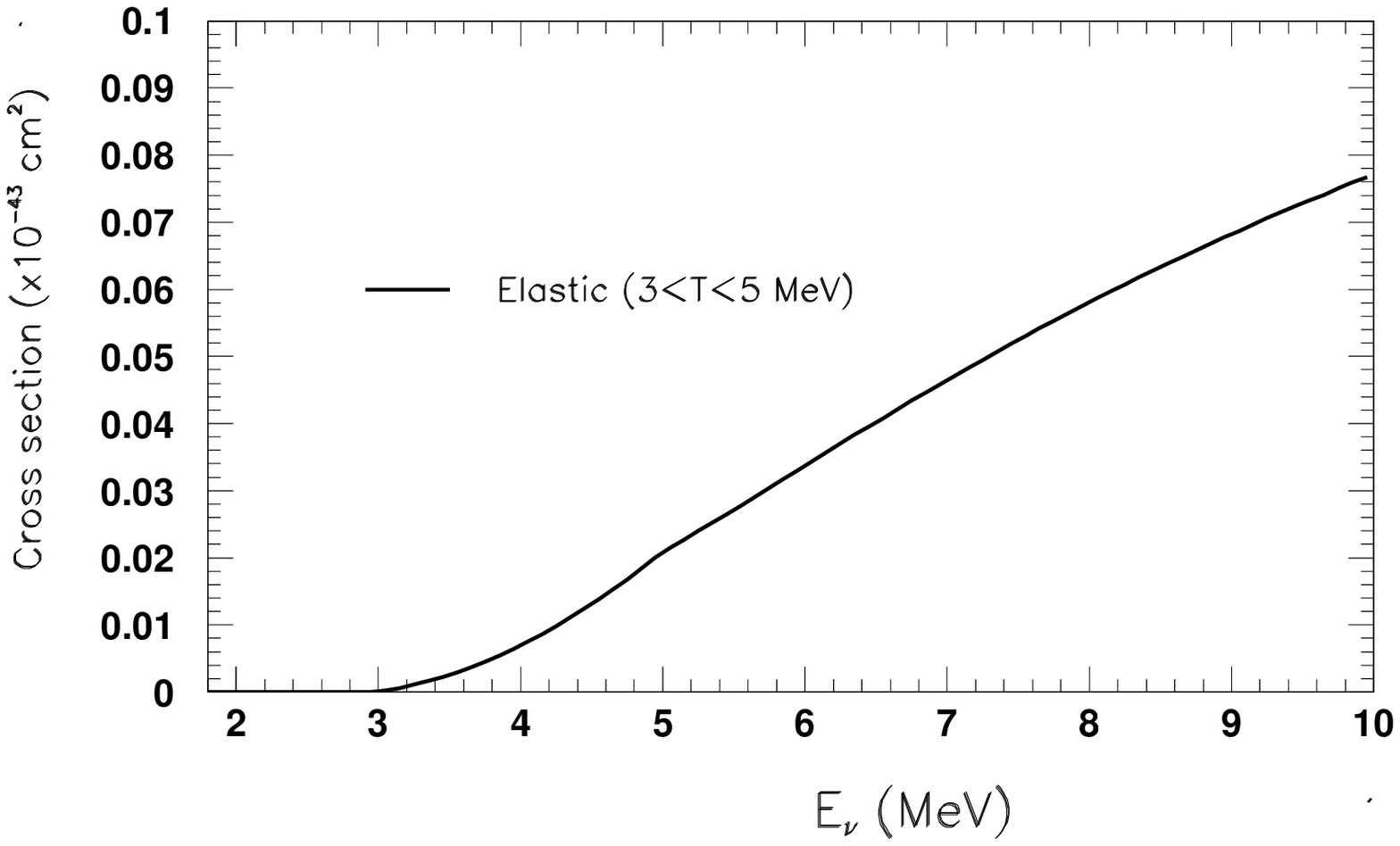}
\includegraphics[width=7.5cm,bb=70 370 544 686, clip=true]{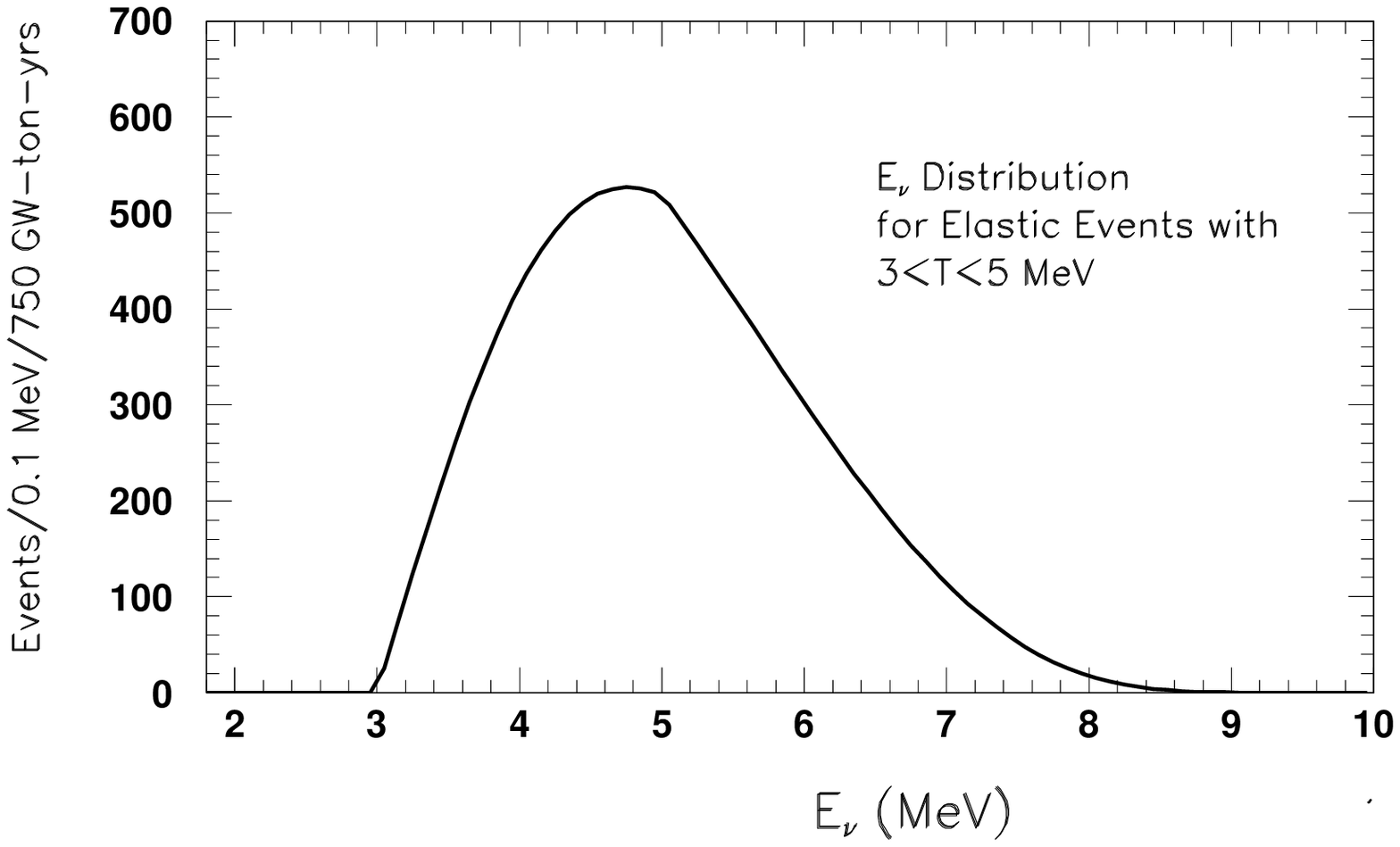}
\includegraphics[width=7.5cm,bb=70 370 544 686, clip=true]{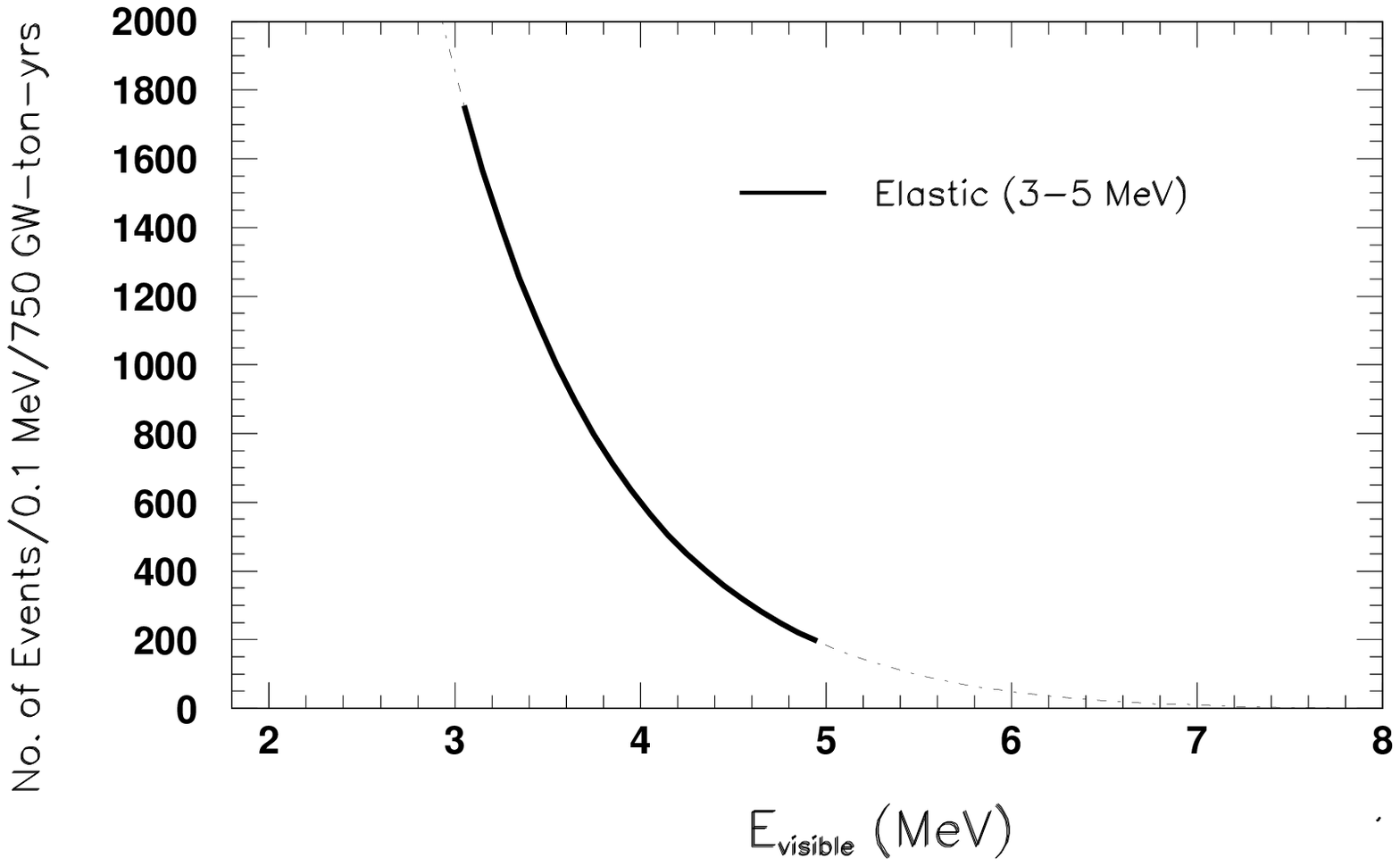}
\vspace{1mm}
\caption{\label{getESevents}Illustration of method for comparing elastic scattering interaction prediction to data. Top: elastic scattering cross section for interactions with visible energy between 3 and 5 MeV, for arbitrary $\sin^{2}\theta _{w}$, shown as a function of neutrino energy. Middle: resulting
predicted event rate when the cross section (top) is multiplied by the flux
distribution (Fig.~\ref{getibdflux}, bottom). Bottom, bold: Rebinning of predicted event rate as a function of $E_{vis}$. Dotted line shows how the prediction would extend beyond the energy window. This predicted distribution will be compared with data.}
\end{figure}

The $\sin^{2}\theta_{W}$ measurement requires that the absolute
$\overline {\nu}_{e}$ flux be known with good accuracy. The flux can
be measured in a straightforward manner using the high statistics
sample of $\overline{\nu}p$ inverse beta-decay events.   This process 
is illustrated in Fig.~\ref{getibdflux}.  There is a one-to-one correspondence
between visible energy and neutrino energy for these events:
\[
E_{\nu}=E_{vis}+1.8-2m_{e}~.%
\]
As shown in Fig.~\ref{getibdflux} (top), therefore, the events can be binned
as a function of $E_\nu$.  The $\overline{\nu}p$ inverse beta-decay
cross section, shown in Fig.~\ref{getibdflux} (middle), is very well
known, both in shape and magnitude, from measurements of the neutron
lifetime.  This has an uncertainty of 0.2\%.  As a result, the flux
can be extracted for neutrinos above the threshold energy for inverse
beta decay, see Fig.~\ref{getibdflux} (bottom).  This is the same flux
which contributes to our signal events.

To extract the predicted number of signal events, we use the
procedure illustrated by Fig.~\ref{getESevents}.  The top plot in this
figure shows the cross section for elastic scattering events with 3 to
5 MeV visible energy as a function of $E_\nu$.  Multiplying this cross
section by the flux in Fig.~\ref{getibdflux} (bottom), results in a
total number of elastic scattering events with the visible energy
cut, binned as a function of true $E_\nu$.  This distribution is shown in
Fig.~\ref{getESevents} (middle).  To see this, we rebin these events
according to $E_{vis}$ and we obtain Fig.~\ref{getESevents} (bottom).
This distribution is the prediction which will be compared with
data.  

We will then vary $\sin^2 \theta_W$ in the cross section,
Fig.~\ref{getESevents} (top), to obtain the best agreement between
data and prediction.  While the sensitivity is expected to mainly rely
on normalization, the shape comparison will provide an important cross
check.

\begin{figure}
\vspace{5mm}
\centering
\includegraphics[width=8.5cm,bb=72 370 540 654, clip=true]{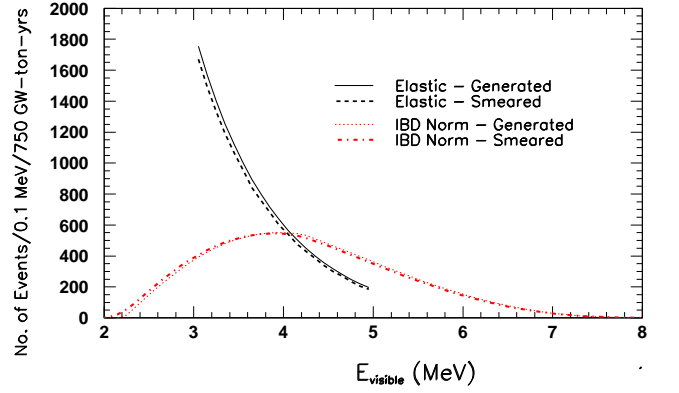}
\vspace{1mm}
\caption{\label{Evis_Norm}Comparison of generated and smeared visible energy distributions.  Solid, Black: elastic scattering, generated; Dashed, Black:
elastic scattering, smeared; Dotted, Red: inverse beta-decay generated;
Dot-dashed, Red: inverse beta-decay smeared. The inverse beta-decay events
(IBD Norm) have been weighted by the elastic to inverse beta-decay cross
section ratio.}
\end{figure}

The error in the $\overline{\nu} e$ event prediction as a function
of $\sin^{2}\theta_{W}$ has contributions from statistical and
systematic uncertainties. The statistical uncertainty is related to
the $\overline{\nu}p$ inverse beta-decay event sample used to
determine the flux. For the assumed ``generic experiment'' described
here, there are about $2.7\!\times\!10^{6}$ $\overline{\nu}p$ inverse
beta-decay events which, when weighted by the cross section ratio of
$\overline{\nu}_{e}e$ to $\overline{\nu}p$ interactions and Gd capture
fraction, yields an effective number of $1.58\!\times\!10^{6}$ events. 
As a result, the statistical
error associated with the flux normalization and energy dependence is
very small due to the high statistics, giving a contribution of 0.08\%.

\begin{figure}
\vspace{5mm}
\centering
\includegraphics[width=8.5cm,bb=80 370 562 654, clip=true]{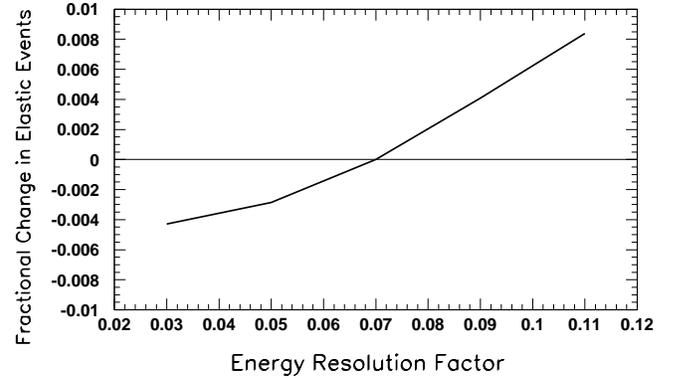}
\vspace{1mm}
\caption{\label{Evis_k}Change in number of elastic scattering events versus energy resolution factor.  }
\end{figure}

The elastic scattering events have a substantially different visible
energy distribution, as compared to the inverse beta-decay events
after weighting by the elastic to inverse beta-decay cross section
ratio (Fig.~\ref{Evis_Norm}).  The reconstructed
energy resolution smearing will therefore affect the two distributions
differently, and a correction will need to be applied when using the
inverse beta decay events for the elastic event prediction.  Assuming
an energy resolution of $\Delta E/E = 0.07/\sqrt{E(MeV)}$, the number
of elastic scattering events in the $3<E_{Visible}<5$ MeV region goes
down by 5\%, but the sum over the full region of the weighted inverse
beta decay events changes very little, as seen in
Fig.~\ref{Evis_Norm}.  One therefore needs to make a correction using
a Monte Carlo simulation of the smearing and apply it to the
prediction.  This correction will depend on knowing the energy
resolution for the detector.  Fig.~\ref{Evis_k} shows the fractional
change in the predicted number of elastic scattering events versus the
energy resolution factor $k$ used in the parameterization, $\Delta E/E
= k/\sqrt{E(MeV)}$.  Sources and other types of calibrations will be
used to determine $k$ for the experiment.  For now, it is assumed that
$k=0.07$ with an uncertainty of 10\% or $\pm 0.007$.  As seen from
Fig.~\ref{Evis_k}, this gives a systematic error on the normalization
of 0.1\% due to the uncertainty in $k$.  This is negligible for the
analysis.

Similarly, but more importantly, an energy scale error or an energy 
offset error will also affect the elastic scattering and inverse beta 
decay samples differently.  Thus this systematic error does not effectively
cancel.   There are a number of ways to constrain energy scale and offset 
errors.  The best method uses the large sample of $^{12}$B decays,
which are $\beta$ decay events.  If one can justify the conversion from
photon to electron response, then one can also use the millions of 
reconstructed 2.2 MeV photons from neutron capture on Hydrogen. 
The peak at 4.9  MeV from capture on Carbon also offers a useful sample.
In this paper, we assume a systematic error on the elastic scattering rate of 0.5\% from energy scale
or offset.  This corresponds to knowledge of the energy scale to 0.33\%
or the offset to 7.5 keV.   These are ambitious goals and clearly 
calibration is a high priority for the analysis.

There are important systematic errors associated with the 
number of free protons in the target and the number of electrons available as targets for elastic scattering.  These systematic errors are correlated and the correlations need to be taken into account in calculating the normalization uncertainty.  With this procedure
the fractional error on the number of electrons is 75\% of the error
on the number of free protons.  CHOOZ determined the number of free
protons by burning their target material~\cite{CHOOZfinal}, yielding a measurement
accurate to 0.8\%.  Assuming we can do no better, the fractional error
on the number of electrons is, therefore, 0.6\%.

Another important systematic is the error on the fraction of neutrons
which will be tagged by a Gd capture.  As discussed in
Sec.~\ref{introdesign}, we use this sample because it selects
events with negligible background.  For a systematic error estimate,
we assume that $84.00\pm 0.25$\% events will capture on Gd.

\section{\label{errors}Calculating the Error on $\sin^2 \theta_W$}

We first obtain the error on the number of signal events and then
extract the error on $\sin^2 \theta_W$.  The terms which contribute to
the error on the number of signal events are: 1) the statistical error
on the signal; 2) the statistical and systematic errors associated
with the $\overline{\nu} p$ background; 3) the statistical and systematic
errors associated with the environmental backgrounds; and 4) the
statistical and systematic errors associated with the normalization.

For the first calculation, we assume the standard set of proposed
cuts.  Next, we consider what is required to reach the NuTeV level of
error.  Then, we consider the impact if the experiment has 
less scintillator purity than proposed here.    Lastly, we consider the
impact if there is substantially more background from isotope decays
than expected.

\subsection{\label{proposederr}Error on $\sin^2 \theta_W$ for the Proposed Analysis}

\begin{table}[tbp]
\begin{center}
{
\begin{tabular}{|c|c|} \hline
Statistical error on the signal & 0.95\% \\
Statistical error $\overline{\nu} p$  background subtraction &  0.43\%\\
Systematic error $\overline{\nu} p$  background subtraction &  0.0\%\\
Statistical error on U and Th background & 0.09\% \\
Systematic error on U and Th background & 0.0\% \\
Statistical error on muon-induced isotopes & 0.33\% \\
Systematic error on muon-induced isotopes & 0.20\% \\
Statistical error on the normalization & 0.10\% \\
Systematic error from energy scale/offset & 0.50\% \\
Systematic error on electron-to-free-proton ratio & 0.60\%\\
Systematic error on the Gd capture fraction & 0.30\% \\ \hline
Total error & 1.40\%
\\ \hline
\end{tabular}}
\caption{\label{tab:errsig}Fractional errors contributing to the error on the number of $\overline{\nu} e$ scattering events based on assumptions presented in this paper.  To equal the NuTeV error, 1.2\% total error was required.}
\end{center}
\end{table}

In this section, we consider the contribution of each error source.  A
summary of each of the sources, along with the fractional error on the
number of $\overline{\nu} e$ events, is shown in Table~\ref{tab:errsig}.
Where the contribution to the error was negligible ($<10$ events), we
list 0\% error.

The statistical error on the signal is calculated using the number of
elastic scattering events in the $E_{vis}$ window, $N_{e}$.  We find that 
for 900 days live-time, $N_e=11,400$ events.

The statistical error from the $\overline{\nu} p$ background subtraction is
$\sqrt{N_{p~bkgd}}/N_e$, where $N_{p~bkgd}$ is the number of 
$\overline{\nu} p$ events passing the signal cuts.  For 900 days live-time, 
we expect $N_{p}=2.70\!\times\!10^6$ $\overline{\nu} p$ events with a 
rejection efficiency of which only 2430 survive all cuts.
We assume that the systematic error on the $\overline{\nu} p$ background 
measurement is negligible (see Sec.~\ref{nubarp} for justification).

The environmental backgrounds contribute both statistical and
systematic errors.  The statistical contribution is given by
$\sqrt{N_{env}}$, and the contribution from the systematic error is 
$\sigma_{env~sys}/N_e$.  We list the contribution from U and Th and
muon-induced isotopes separately in Table \ref{tab:errsig}.  The
systematic error on the $^{238}$U and $^{232}$Th contaminants is 
negligible but the systematic error on the muon-induced isotopes, 
$\pm$23 events, must be considered.

The statistical error due to the size of the normalization sample is 
very small, $\sim$0.08\%, as described in Sec.~\ref{norm}.  The first 
systematic  error on the normalization comes from the uncertainty in the 
number of electrons in the sample, which is tied to the uncertainty in
the number of free protons and is $dN_{target\ electrons}/N=0.6$\%.  The second
significant systematic error on the normalization sample comes from
the error on the knowledge of the Gd capture fraction.  We have argued
that 0.25\% can be attained.

Adding the systematic errors in quadrature, we find
$(dN/N)_{sys}=1.03$\%.  Adding this in quadrature with the statistical
error on the signal yields $(dN/N)_{tot}=1.40$\%.  This is close to the
goal of 1.15\% which we set at the start of this paper.  Extracting the
error on $\sin^2 \theta_W$, we obtain: $\delta(\sin^2 \theta_W)=0.0020$.
This is comparable to the NuTeV error of 0.00164.

\subsection{Improving this Measurement}

As one can see from Table~\ref{tab:errsig}, the error on $\sin^2
\theta_W$ is dominated by statistics.  If the running period were 
doubled to 1800
days, one would achieve $\delta(\sin^2 \theta_W)=0.0018$.  Alternatives
to increased running period include enlarging the detector, adding 
extra near detectors, finding a closer approach to the reactor cores,
or moving to a more powerful reactor.   Any of these options
will incrementally improve the result.

The rate of production of muon-induced isotopes can be reduced by
using a larger overburden.  If the rate reduced by a factor of two by
going to 450 mwe, then the error on $\sin^2 \theta_W$ drops to
0.0019.  If we have a 450 mwe overburden and the experiment runs for
1800 days, the experiment attains $\delta(\sin^2 \theta_W)=0.0017$.
Reduction of these background events may also be achieved by a better
muon-neutron veto.  However, if the veto introduces excessive
deadtime the loss in elastic scattering statistics may offset
the gains in background reduction.

\subsection{\label{baddope}Impact of Impurities of the Gd-Dopant}

We have assumed that KamLAND levels of purity for U and Th ($5\!\times\!
10^{-17}$ g/g Th) can be achieved.  The techniques for purifying oil
have been established by CTF and KamLAND and therefore appear to be
practical.  This experiment, however, requires that Gd dopant be added
to the oil.  Experience from CHOOZ~\cite{CHOOZfinal} indicates that
this can introduce a high level of Th contamination, so
purification of the Gd will need to be researched.

To establish the effect of increased contamination, consider the
change in the error as the contamination is increased.  In our
calculation above, with 100 Th and U induced events in the $E_{vis}$
window, we achieved $\delta(\sin^2 \theta_W)=0.0020$.  Two orders of 
magnitude increase in U and Th contamination gives $\delta(\sin^2 \theta_W)=0.0024$, which is within a tolerable range, especially given 
that running longer will still result in a substantial reduction in error.  
However, three orders of magnitude larger contamination renders the result
uninteresting.  This constrains the necessary level of purity which must 
be achieved to better than $5\!\times\!10^{-15}$ g/g of Th.

\subsection{Impact of Increased Isotope Background}

We have calculated the background from $\beta$-decaying isotopes based
on calculations from reference~\cite{Hagner}.    We did not, however,
include the background from $^{11}$Be, for which only a gross upper 
limit has been set~\cite{Hagner}.   Using this upper limit as 
an expected level of isotope production increases the background from
1458 to 1803 events.   This produces a negligible shift in 
$\delta(\sin^2 \theta_W)$.  In fact, increasing the total isotope 
background by a factor of two only increases $\delta(\sin^2 \theta_W)$
to 0.0021.

\section{\label{conclusions}Conclusions}

This paper discusses a technique for measuring $\sin^2 \theta_W$ at a
reactor-based experiment using $\overline{\nu} e$ elastic scatters.  A
precise measurement of $\sin^2 \theta_W$ at $Q^2\approx 4\times 10^{-6}$~GeV$^2$, 
neutrinos as probes opens a window for tests of neutrino properties
and electroweak theory.  We have used an experimental design which
is consistent with many proposals for near detectors at reactor-based
oscillation experiments.  We have also assumed realistic reactor power
and a human-scale run time of about 900 days.

The analysis has statistical and systematic errors which are roughly
equal, so increased statistics will yield further improvement.  At
least $\sim$26 tons of fiducial volume are required and the detector
should be located as close as possible to the reactor ($\sim$250~m or
less).

We have also introduced the idea of normalizing to the $\overline{\nu} p$
events.  This substantially reduces the error from the flux. 
Because the normalization sample is measured in the same detector as
the elastic scattering signal, many systematics effectively cancel,
including those associated with deadtime and fiducial volume.

We have also considered backgrounds from misidentified inverse beta
decay events.  Using $n$-identification and a visible energy window,
this background can be reduced to an acceptable level.  Environmental
background is dominated by the contribution from spallation by cosmic
ray muons which produce isotopes which $\beta$-decay.  To attain
acceptable rates, 300 mwe overburden is the minimum required.  Indeed,
deeper overburden is desirable.  Future studies on reducing this
background by identifying muons which cause spallation are important.  It is
also necessary to maintain oil and Gd purity from U and Th
contaminants.  Our calculations show that KamLAND-levels of purity are
desirable, but an increase of two orders of magnitude of impurity is
acceptable.

This exercise was meant to serve as a proof-of-principle that a
reasonable error on $\sin^2 \theta_W$ can be attained at a reactor-based 
experiment.  The technique has not yet been fully optimized.  The total 
error which we obtain on $\sin^2 \theta_W$ is $\delta(\sin^2 
\theta_W)$=0.0020.  This is similar to the NuTeV error of 0.00164 and is 
lower than the published SLAC E158 and APV results.  A measurement at this precision will uniquely probe the electroweak and neutrino sectors and is 
currently being explored by several groups~\cite{Rosner:2004yt,Fisher}.  
Based on this study, we conclude that the idea is feasible and more 
detailed studies are warranted.

\subsection*{Acknowledgments}
We thank G.~P.~Zeller, H. Newfield-Plunkett and the Columbia Neutrino Group
for their input.    Also, we thank 
S.~Biller, T.~Bolton, J.~Formaggio,  K.~Heeger, P.~Fisher, G.~Gratta, R.~Imlay, W.~Louis,
D.~Naples, P.~Nienaber and P.~Vogel for their comments.

\bibliography{sin2thw}

\end{document}